\documentclass[12pt]{iopart}
\usepackage{graphicx}
  \expandafter\let\csname equation*\endcsname\relax
  \expandafter\let\csname endequation*\endcsname\relax
\usepackage{amsmath,amssymb,bm}
\usepackage{amsfonts}
\usepackage{multirow}
\usepackage{algorithm}
\usepackage{rotating}
\usepackage{hyperref}
\usepackage[noend]{algpseudocode}

\usepackage[color=green!40]{todonotes}

\newcommand{\boldtau}{\mbox{\boldmath $\tau$}}
\newcommand{\boldsigma}{\mbox{\boldmath $\sigma$}}

\begin{document}
\title{Bayesian optimization in \textit{ab initio} nuclear physics}
\author{A. Ekstr\"{o}m$^1$, C. Forss\'en$^1$,
  C. Dimitrakakis$^{2}$, D. Dubhashi$^2$,\\ H. T. Johansson$^1$,
  A. S. Muhammad$^2$, H. Salomonsson$^2$, A. Schliep$^3$}
\address{$^1$Department of Physics, Chalmers University of Technology,
  SE-412 96 Gothenburg, Sweden }
\address{$^2$Department of Computer
  Science and Engineering, Chalmers University of Technology, SE-412
  96 Gothenburg, Sweden}
\address{$^3$Department of Computer Science and Engineering, Chalmers University of Technology and University of Gothenburg, SE-412 96 Gothenburg, Sweden}
\begin{abstract}
Theoretical models of the strong nuclear interaction contain unknown
coupling constants (parameters) that must be determined using a pool
of calibration data. In cases where the models are complex, leading to
time consuming calculations, it is particularly challenging to
systematically search the corresponding parameter domain for the best
fit to the data. In this paper, we explore the prospect of applying
Bayesian optimization to constrain the coupling constants in chiral
effective field theory descriptions of the nuclear interaction. We
find that Bayesian optimization performs rather well with
low-dimensional parameter domains and foresee that it can be
particularly useful for optimization of a smaller set of coupling
constants. A specific example could be the determination of leading
three-nucleon forces using data from finite nuclei or three-nucleon
scattering experiments.

\end{abstract}
\maketitle
\section{Introduction}
Mathematical optimization plays a central role in natural
science. Indeed, most theoretical predictions are preceded by a
calibration stage whereby the parameters of the model are optimized to
reproduce a selected set of calibration data. In nuclear physics, the
coupling constants of any theory of the strong interaction between
protons and neutrons (nucleons) must be determined from experimental
data before one can attempt to solve the Schr\"{o}dinger equation to
make quantitative predictions of the properties of atomic nuclei.

Typically, measured low-energy nucleon-nucleon ($NN$) cross sections
and the properties of light nuclei with mass number $A \leq 4$ have
been used for calibrating the $NN$ and three-nucleon ($NNN$)
interaction sectors of the nuclear force, see
e.g. Refs.~\cite{Stoks,Wiringa:1994wb,Gazit:2008ma} and references
therein. However, it is still an open question---with several parallel
and ongoing research efforts~\cite{Reinert, Carlsson:2016,
  Piarulli,Elhatisari:2016dv}---how to best constrain the unknown
coupling constants in theoretical descriptions of the nuclear
interaction and how to incorporate the covariance structure of the
experimental data and the model
discrepancy~\cite{Carlsson:2016,Melendez,PhysRevLett.115.122301,Wesolowski}.
Recent theoretical studies~\cite{
  Elhatisari:2016dv,Ekstrom:2015gwa,Lapoux:2016ju, Hagen:2016io}
indicate that $NN$ scattering data and the properties of very light
nuclei do not contain enough information to constrain all directions
in the parameter domain at sufficient level. Instead, it has been
proposed that the pool of fit data need to be augmented with
complimentary data types, such as $NNN$ scattering cross sections,
and/or the ground-state properties of nuclei with $A>4$, or even
empirical properties of infinite nuclear matter. However, modeling of
such observables is typically much more complex and requires substantial
computational effort ranging from hours to days for just one model
evaluation, even on a supercomputer. Consequently, the optimization of
the underlying model parameters will be difficult. The main focus of
the present work is to investigate a possible strategy for mitigating
this computational challenge.

Inspired by recent progress in the optimization of hyperparameters of
deep neural networks~\cite{Snoek:2012}, we explore several Bayesian
optimization (BayesOpt) strategies\footnote{we
employ the BaysOpt implementation provided through the Python package
GPyOpt~\cite{gpyopt2016}.} for maximizing the likelihood of
objective functions based on complex models in nuclear
physics. BayesOpt originated more than 50 years ago~\cite{Kushner}, it
was popularized in the 1990s, see e.g.~\cite{Mockus1998,Jones}, and
has since then been applied in various disciplines; from selecting
experiments in material and drug
design~\cite{doi:10.1287/ijoc.1100.0417} to tuning event-generators in
particle physics~\cite{1748-0221-12-04-P04028}. The central idea in
BayesOpt is to construct a probabilistic surrogate model, usually a
Gaussian process ($\mathcal{GP}$), to capture our prior beliefs and
knowledge about the objective function, $f(\mathbf{x})$, and
iteratively confront the surrogate with actual data samples from
$f(\mathbf{x})$ and thereby refine our posterior of this function. The
main advantage of BayesOpt is that we can incorporate prior beliefs in
a straightforward way. The disadvantage lies in the arbitrariness and
uncertainty of \textit{a priori} information.

In the following we will be dealing with complex models in nuclear
physics. The origin of the underlying physics model and its parameters
is briefly introduced in this section, while more details and
relevant references are provided in~\ref{app:nnlo}.
Nucleons are made of quarks and gluons, and it is well known that the
strong interaction between these fundamental particles is described in
detail by quantum chromodynamics (QCD), which is part of the standard
model of particle physics. It is equally well-known that
QCD is not perturbative in the low-energy region relevant for nuclear
structure physics. This prevents straightforward application of,
e.g., perturbation theory to compute atomic nuclei starting from QCD. 
Instead, chiral effective field theory
($\chi$EFT)~\cite{Bedaque:2002gm, EHM09, 0000000102130473} is
constructed as a low-energy approximation to QCD. This framework shows
promising signs of being an operational approach to analyzing atomic
nuclei while maintaining a firm link with the more fundamental
theory. In $\chi$EFT, the long-ranged part of the nuclear interaction
is described in terms of pion exchanges, while the short-ranged part
is parameterized by contact interactions. The unknown parameters of
this description are known as low-energy constants (LECs). It is
important to realize that any realistic description of the nuclear
interaction has to introduce unknown coefficients, but it has been
found that $\chi$EFT is able to capture the physics with a relatively
small number of parameters. Depending on the level of details
included, there are typically $\sim 10$--$30$ LECs.

Clearly, if model predictions of physical observables, such as
scattering cross sections, are computationally expensive, we face the
challenge of optimization with a limited budget of evaluations of the
objective function. For the overwhelming majority of well-developed
research problems there does not exist a universal optimization
strategy that guarantees to arrive at a global optimum of an objective
function in a finite number of iterations. Instead, any information
regarding the mathematical or computational structure of the objective
function, perhaps guided by the physical nature of the underlying
problem, should play an important role in the choice or design of the
optimization algorithm.

In this paper, we will systematically study the application of
BayesOpt to optimization problems of increasing degree of
complexity. The BayesOpt algorithm is presented in
Sec.~\ref{sec:bayesopt} with its main ingredients:
the $\mathcal{GP}$ kernel and the acquisition function. 
The performance of different optimization algorithms can be compared
using a data profile. This measure, as used in the present work, is
introduced in Sec.~\ref{sec:performance}. 
In order to benchmark the performance of BayesOpt with various
settings in controlled problem conditions we will employ a selected
set of six test functions in $D=2,4,8$ dimensional parameter
domains. This study is presented in Sec.~\ref{sec:test_functions},
while the test functions themselves are listed
in~\ref{app:testfunctions}.
The main focus of this work is found in Sec.~\ref{sec:NP} with the
application of BayesOpt to a real nuclear physics problem. We will use
BayesOpt to optimize the 12 LECs appearing at next-to-next-to-leading
order (NNLO) in $\chi$EFT, using the proton-neutron scattering data in
the Granada database~\cite{Navarro13_2} with laboratory energies of
the incident proton beam below 75 MeV. This case is complex enough to
constitute a non-trivial problem from a physics as well as an
optimization perspective. However, it is still computationally
straightforward such that we can easily afford a detailed analysis of
12 different BayesOpt algorithms. Indeed, each evaluation of the
objective function at a specific point in the parameter domain only
takes a couple of seconds on a standard desktop computer. Still,
evaluating the 4,096 corners in the hypercube of the corresponding
parameter domain will take a couple of hours, so the premise for
BayesOpt is well justified. We will compare the optimization
performance of BayesOpt with the POUNDERs algorithm~\cite{SWCHAP14}.
This is a simulation-based optimization algorithm that has been
successfully applied in non-linear least-squares optimization in
nuclear physics before~\cite{Kortelainen2010, Ekstrom2013}.
Finally, we end with a summary and outlook in Sec.~\ref{sec:summary}.

\section{Bayesian optimization \label{sec:bayesopt}}
\begin{figure}
  \begin{center}
    \includegraphics{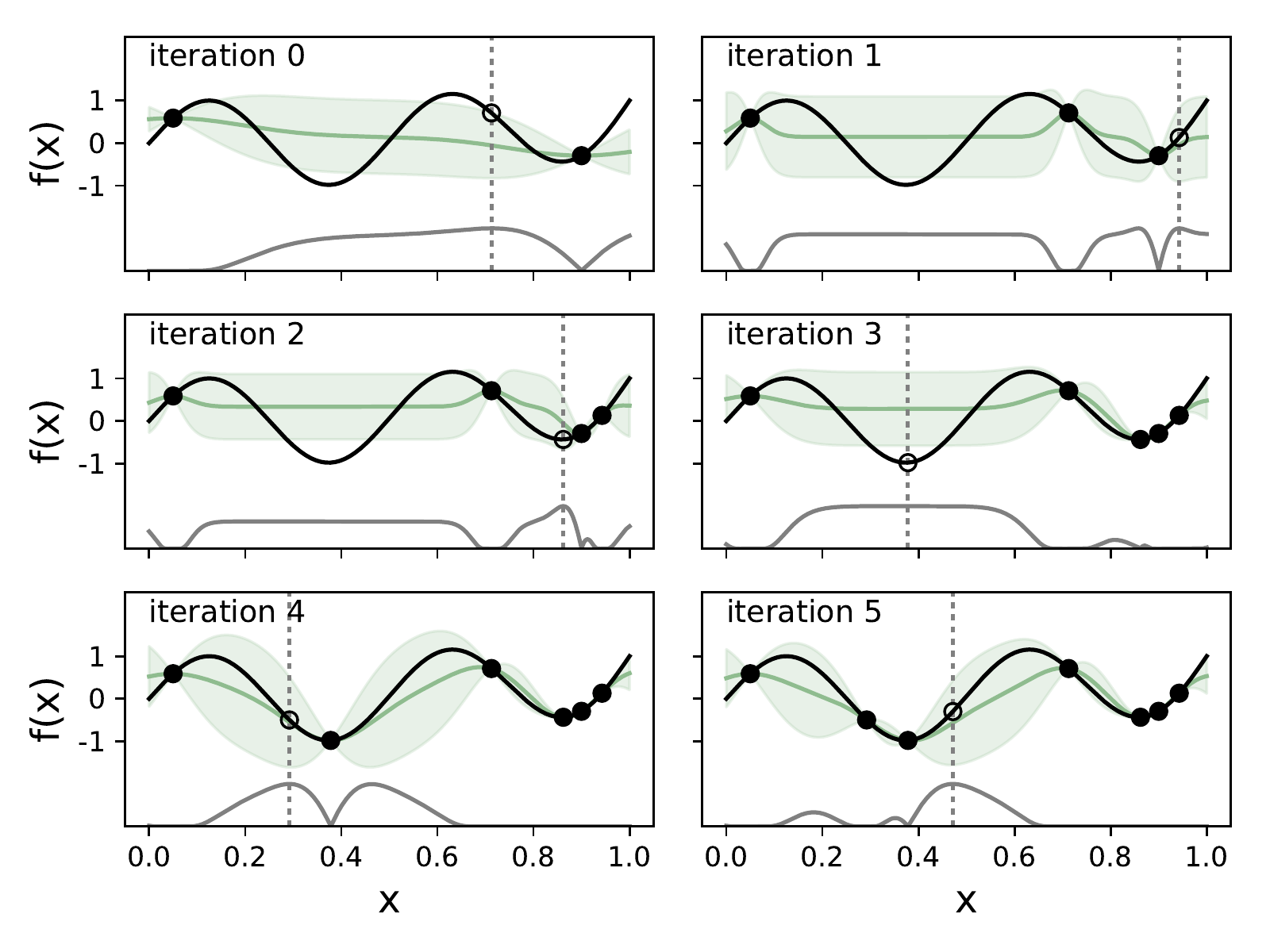}
    \caption{Five BayesOpt iterations towards finding the global
      minimum of the function $f(x)=\sin(4\pi x) + x^{4}$ where
      $x\in[0,1]$ (black solid line) using the expected improvement
      acquisition function (bottom gray line). Two initial function
      evaluations $f(x=0.05)$ and $f(x=0.9)$ (filled markers) were
      randomly selected. In each iteration the \textit{next}
      evaluation point (dashed gray line) in the parameter domain is
      determined from the maximum of the expected improvement
      acquisition function. For each iteration, the corresponding
      value of $f(x)$ is indicated with an empty marker. The mean
      (green line) and 95\% confidence interval (green region) of a
      $\mathcal{GP}$ with a squared-exponential kernel, sequentially
      approach $f(x)$. After iteration 3, the algorithm leaves the
      rightmost $x$-domain, and the associated local minimum, to
      explore the region containing the global minimum.}
    \label{fig:bayesopt_tutorial}
  \end{center}
\end{figure}

Without any loss of generality, we will frame the determination of the
LECs in $\chi$EFT as a minimization problem. Global minimization of a
function $f:\mathbb{R}^{D} \rightarrow \mathbb{R}$, with input
parameters $\mathbf{x}$ that are perhaps subject to some constraints
$c(\mathbf{x})\leq 0$ and typically belong to a compact input domain
$\mathcal{X} \subset \mathbb{R}^{D}$, is a long-standing and central
problem in science. Here, we also specialize the formalism to scalar
valued functions $f$. Mathematically, we are trying to find a global
minimizer:
\begin{equation}
  \mathbf{x}_{\star} = \underset{\mathbf{x} \in \mathcal{X}}{\textnormal{arg min}}, \, f(\mathbf{x}),
\end{equation}
i.e. a point that fulfill $f(\mathbf{x}_{\star}) \leq f(\mathbf{x})$
for all $\mathbf{x} \in \mathcal{X}$. In general, this is an
intractable problem unless we have detailed information about
$\mathbf{x}$ or that the parameter domain only contains a finite
number of points. In reality, we are trying to find \textit{local}
minimizers to $f(\mathbf{x})$, i.e. points $\mathbf{x}_{\star}$ for
which $f(\mathbf{x}_{\star}) \leq f(\mathbf{x})$ for all $\mathbf{x}
\in \mathcal{X}$ close to $\mathbf{x}_{\star}$. A substantial
complication arises if $f(\mathbf{x})$ is computationally expensive to
evaluate. Then it becomes even more important to adopt a well-founded
strategy, based on all present knowledge about $f(\mathbf{x})$, for
carefully selecting a dataset of $n$ sequential function queries
$\mathcal{D}_{1:n} = \{\mathbf{x}_{i},y_{i}\}_{i=1}^{n}$, where
$y_i=f(\mathbf{x}_{i})$, such that we increase our chances of a rapid
convergence towards $\mathbf{x}_{\star}$. A BayesOpt framework can
provide this. Moreover, at each iteration BayesOpt will consider all
insofar collected data points and thereby take full advantage of the
history of the optimization run. Note that we refer to a set of
function evaluations as \textit{data}. This should not be confused
with experimental data. There are two main components in a BayesOpt
algorithm;
\begin{itemize}
  \item A prior probabilistic belief $p(f| \mathcal{D})$ for the
    function $f$ given some data $\mathcal{D}$. The prior is often a
    $\mathcal{GP}$. This is updated in every iteration.
  \item An acquisition function $\mathcal{A}(\mathbf{x}| \mathcal{D})$
    given some data $\mathcal{D}$, i.e. a heuristic that balances
    exploration against exploitation and determines where to evaluate
    the objective function $f(\mathbf{x})$ next.
  \end{itemize}
The next iterate, $\mathbf{x}_{i+1}$, is selected where we expect the
minimizer $\mathbf{x}_{\star}$, based on some utility function. Below,
we will define two different acquisition functions
$\mathcal{A}(\mathbf{x}|\mathcal{D})$, and show how to embed them in
an iterative context for selecting sample points $\mathbf{x}_{i}$. In
the following we will also drop the explicit data dependence in the
notation for the acquisition function and only write
$\mathcal{A}(\mathbf{x})$. A pseudo-code for BayesOpt is listed in
Algorithm~\ref{BOalgo} and a pictorial exposition of a handful of
BayesOpt iterations of a simple univariate function is provided in
Fig.~\ref{fig:bayesopt_tutorial}.

\begin{algorithm}
  \caption{Bayesian Optimization}\label{BOalgo}
  \begin{algorithmic}[1]
    \State select initial $\mathbf{x}_{1},\mathbf{x}_{2},\ldots \mathbf{x}_{k}$, where $k \geq 2$
    \State evaluate the objective function $f(\mathbf{x})$ to obtain $y_i=f(\mathbf{x}_i)$ for $i=1,\ldots,k$
    \State initialize a data vector $\mathcal{D}_k = \left\{(\mathbf{x}_1,y_1),(\mathbf{x}_2,y_2),\ldots, (\mathbf{x}_k,y_k) \right\}$
    \State select a statistical model for $f(\mathbf{x})$
    \For{$n=k+1,k+2,\ldots$}
    \State select $\mathbf{x}_{n}$ by optimizing the acquisition function
    \Statex \hspace*{3cm} $\mathbf{x}_{n} = \underset{\mathbf{x}}{\textnormal{arg max}}\, \mathcal{A}(\mathbf{x}|\mathcal{D}_{n-1})$
    \State evaluate the objective function to obtain $y_n=f(\mathbf{x}_n)$
    \State augment the data vector $\mathcal{D}_n = \left\{\mathcal{D}_{n-1} , (\mathbf{x}_n,y_n)\right\}$
    \State update the statistical model for $f(\mathbf{x})$
    \EndFor
    \State \textbf{end for} 
  \end{algorithmic}
\end{algorithm}

\subsection{The prior: a Gaussian process}
To model the prior $p(f|\mathcal{D})$ for the objective function we
use a Gaussian process $\mathcal{GP}(\mathbf{x})$ with mean function
$\mu(\mathbf{x})$ and covariance matrix $\mathbf{K}$ with entries
$k_{ij} = k(\mathbf{x}_i,\mathbf{x}_j)$. The mean and covariance
functions fulfill expected relations; $\mu(\mathbf{x}) =
\mathbb{E}[\mathbf{x}]$ and $k(\mathbf{x},\mathbf{x}') =
\mathbb{E}[(\mathbf{x}-\mu(\mathbf{x}))(\mathbf{x}'-\mu(\mathbf{x}'))]$
for all $(\mathbf{x},\mathbf{x}') \in \mathcal{X}$. Any real-valued
function $\mu(\cdot)$ is permissible, but for $k(\cdot,\cdot)$ the
corresponding covariance matrix $\mathbf{K}$ must be positive
semidefinite. A $\mathcal{GP}$ is one example of a stochastic process
that is very useful in statistical modeling~\cite{Rasmussen}. In
brief, it is a collection of function evaluations $y_{1:n}$ at
$\mathbf{x}_{1:n}$, with mean $\mu_0$ (often shifted to zero), that
are jointly Gaussian, i.e.
  \begin{equation}
    \left[ \begin{array}{c}
      y_1 \\
      \vdots \\
      y_n
    \end{array}\right] \sim \mathcal{N} \left(\left[
    \begin{array}{c}
      \mu_0 (\mathbf{x}_1)\\
      \vdots \\
      \mu_0 (\mathbf{x}_n)
    \end{array} \right],\mathbf{K} = \left[ \begin{array}{ccc}
        k(\mathbf{x}_1,\mathbf{x}_1) & \cdots & k(\mathbf{x}_1,\mathbf{x}_n) \\
        \vdots                       & \ddots & \vdots \\
        k(\mathbf{x}_n,\mathbf{x}_1) & \cdots & k(\mathbf{x}_n,\mathbf{x}_n) 
      \end{array}
      \right]\right), 
  \end{equation}
  where $y \sim \mathcal{N}(\mu,\sigma^2)$ denotes a normally
  distributed random variable $y$ with mean $\mu$ and covariance
  $\sigma^2$.  After conditioning this prior with some data
  $\mathcal{D}_n = \left\{ (\mathbf{x}_1,y_1), \ldots,
  (\mathbf{x}_n,y_n)\right\}$, with mean $\mu_0$, we obtain the mean
  and covariance of the $\mathcal{GP}$ model for the prior according
  to
\begin{align}
  \label{eq:mean_and_covariance}
   \begin{split}
  {}& \mu(\mathbf{x}) = \mu_0(\mathbf{x}) + \left[
    \begin{array}{c}
      k(\mathbf{x_1},\mathbf{x}) \\
      \vdots\\
      k(\mathbf{x_n},\mathbf{x})
    \end{array} \right]^{T}\left[
    \begin{array}{ccc}
      k(\mathbf{x}_1,\mathbf{x}_1) & \cdots & k(\mathbf{x}_1,\mathbf{x}_n) \\
      \vdots                       & \ddots & \vdots \\
      k(\mathbf{x}_n,\mathbf{x}_1) & \cdots & k(\mathbf{x}_n,\mathbf{x}_n) 
    \end{array}\right]^{-1}
  \left[
    \begin{array}{c}
      y_1 - \mu_0(\mathbf{x}_1)\\
      \vdots \\
      y_n - \mu_0({\mathbf{x}_n})
    \end{array}
    \right]\\
  {}& \sigma(\mathbf{x})^2 = k(\mathbf{x},\mathbf{x}) - \left[
    \begin{array}{c}
      k(\mathbf{x_1},\mathbf{x}) \\
      \vdots\\
      k(\mathbf{x_n},\mathbf{x})
    \end{array} \right]^{T}\left[
    \begin{array}{ccc}
      k(\mathbf{x}_1,\mathbf{x}_1) & \cdots & k(\mathbf{x}_1,\mathbf{x}_n) \\
      \vdots                       & \ddots & \vdots \\
      k(\mathbf{x}_n,\mathbf{x}_1) & \cdots & k(\mathbf{x}_n,\mathbf{x}_n) 
    \end{array}\right]^{-1}
  \left[
    \begin{array}{c}
      k(\mathbf{x_1},\mathbf{x}) \\
      \vdots\\
      k(\mathbf{x_n},\mathbf{x})
    \end{array} \right].
   \end{split}
\end{align}
These explicit expressions follow from the fact that the marginal
distribution of a multivariate Gaussian is also Gaussian.  The mean
and variance of the prior, i.e. the $\mathcal{GP}$, for the
schematic example in Fig.~\ref{fig:bayesopt_tutorial} are indicated
with a green line and a green shaded band, respectively.

  In this work we will consider three common types of
  covariance functions, also referred to as kernels:
 \begin{itemize}
 \item Squared exponential: $k(\mathbf{x},\mathbf{x}') = \theta_{0}^2\exp \left( -\frac{1}{2} r^2 \right)$
 \item Matern 3/2: $k(\mathbf{x},\mathbf{x}') = \theta_0^2(1 + \sqrt{3} r) \exp(- \sqrt{3} r)$
 \item Matern 5/2: $k(\mathbf{x},\mathbf{x}') = \theta_0^2(1+\sqrt{5}r + \frac{5}{3}r^2) \exp \left( -\sqrt{5} r \right)$
  \end{itemize}
  where $r^2 \equiv \sum_{i=1}^{D} \frac{1}{\theta_i^2}(x_i-x_i')^2$
  and $\left\{ \theta_{i} \right\}_{i=0}^{D}$ is a set of
  hyperparameters for the kernel. The correlation length(s) $\ell =
  (\theta_1,\ldots,\theta_{D})$ indicate how far you need to move
  (along a particular direction in the parameter domain) for the
  function values to become uncorrelated.  With automatic relevance
  determination (ARD), we optimize the vector of correlation lengths
  separately in each direction of the function domain $\mathcal{X}
  \subset \mathbb{R}^D$. Without ARD, the kernel hyperparameters are
  isotropic. In this work, we extract the hyperparameters using
  maximum likelihood estimation. 

\begin{figure}
   \begin{center}
    \includegraphics{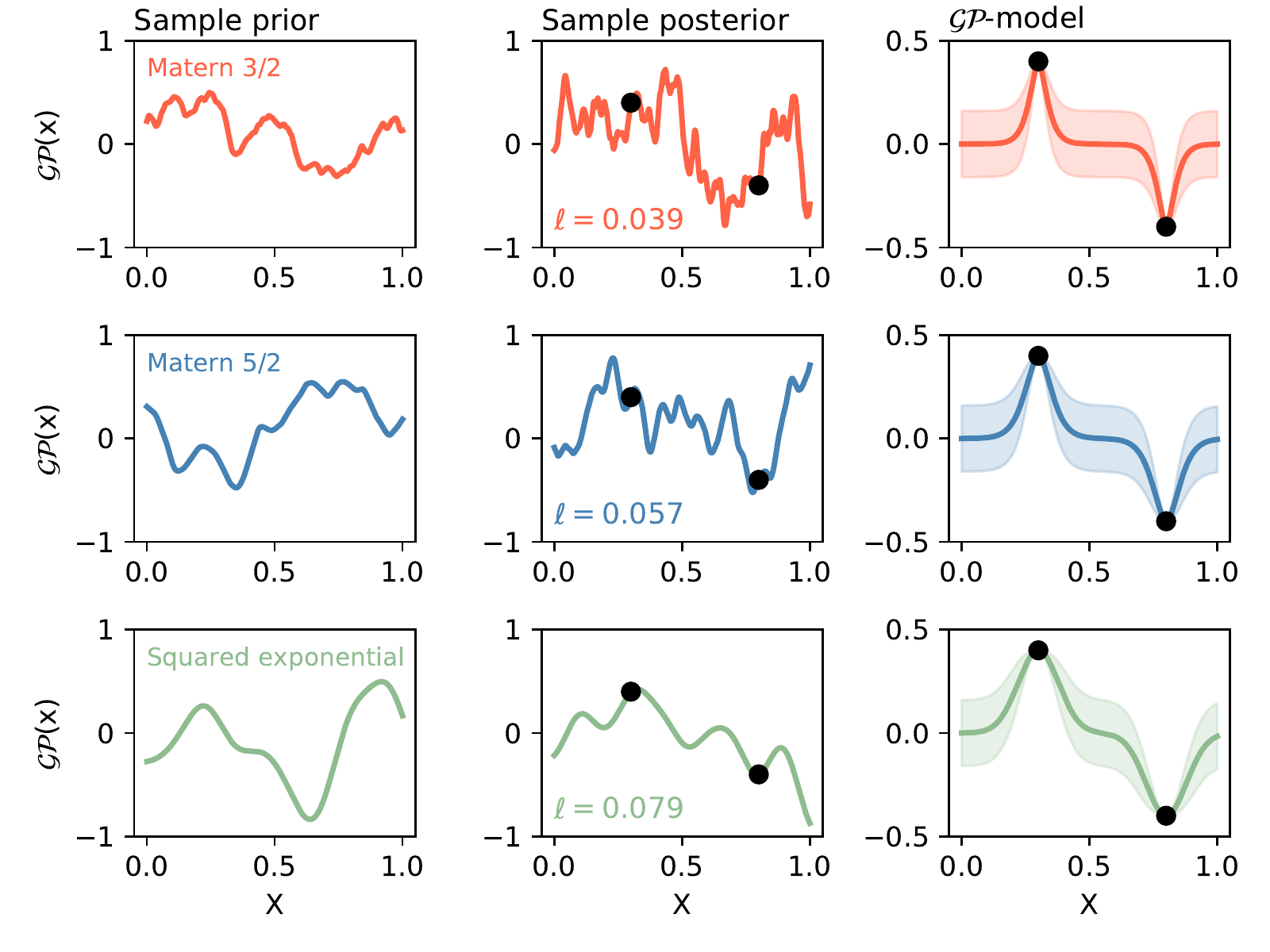}
    \caption{Randomly drawn $\mathcal{GP}$ priors (left column),
      posteriors (middle column), and full probabilistic model (right
      column) using the three kernels we employ in this paper; Matern 3/2
      (first row), Matern 5/2 (middle row), and squared exponential
      (bottom row). For each kernel, the priors are confronted with
      two identical data points (black dots) and the two corresponding
      hyperparameters, covariance and correlation length ($\ell$), are
      optimized to maximize the marginal likelihood of the data. The
      resulting correlation lengths for each kernel are provided in
      the middle column. Clearly, the Matern 3/2, Matern 5/2, and
      squared exponential kernels are increasingly smooth.}
    \label{fig:kernels}
  \end{center}
 \end{figure}

  The characteristic features of the $\mathcal{GP}$s based on the
  three kernels listed above are illustrated in
  Fig.~\ref{fig:kernels}. Their smoothness, equivalent to the typical
  correlation length, is one of the key differences between them. This
  feature of the $\mathcal{GP}$ kernel affects the prior modeling of
  the objective function, and as such the ensuing performance of
  BayesOpt. We will see this clearly in the analyses in
  Secs.~\ref{sec:test_functions}-\ref{sec:NP}.
 
\subsection{The acquisition function}
The acquisition function determines the most likely improvement to the
currently best minimizer in the parameter domain. In
Fig.~\ref{fig:bayesopt_tutorial} the mean of our posterior belief
(green line) of the unknown values of the objective function $f$ (black
line) is sequentially augmented with one new data point (black dot) in
each iteration. The best candidate for further minimizing $f$ at
iteration $n$ is the parameter that maximizes the acquisition function
(red curve). Although the acquisition function is also optimized in
$\mathcal{X} \subset \mathbb{R}^{D}$ it is significantly faster to
evaluate, compared to the underlying objective function
$f(\mathbf{x})$, since it only relies on draws from the prior
$\mathcal{GP}$. Still, the complexity of maximizing
$\mathcal{A}(\mathbf{x})$ increases as we increase the dimensionality
of the parameter domain $\mathcal{X}$. This aspect should not be
underestimated, and it is in fact one of the main challenges with
BayesOpt. Another challenge, although unlikely, could emerge if the
set of collected data points $\mathcal{D}_{1:n}$ becomes very
large. Indeed, in each iteration the evaluation of the $\mathcal{GP}$
requires an inversion of an $n \times n$ matrix, and the complexity of
that operation naively scales as $\mathcal{O}(n^3)$. Cholesky
decomposition reduces this cost somewhat to $\mathcal{O}(n^3/6)$. In
reality, however, this is rarely a limiting factor since we typically
resort to BayesOpt when only a small number of function evaluations
can be carried out in the first place.

A very appealing feature of BayesOpt is its ability to select a new
point $\mathbf{x}_{n}$ in a region of $\mathcal{X}$ where the prior
model of $f$ is exhibiting a large uncertainty. This means that the
algorithm can be rather explorative and therefore escape a local
minimum of the objective function $f$. Depending on the details of
the acquisition function, the explorative nature is balanced with a
certain degree of exploitation, i.e. to evaluate points in the
parameter domain where the prior model for $f$ is exhibiting a low
mean value.  To study the exploration-exploitation balance we will
consider two of the most common acquisition functions; the
\textit{expected improvement} (EI) and the \textit{lower
  confidence-bound} (LCB). In the following, we denote by $f_{\rm min}$
the insofar lowest recorded value of $f(\mathbf{x})$.

The expected improvement acquisition function is defined by the
expectation value of the rectifier ${\rm max}(0,f_{\rm min} -
f(\mathbf{x}))$, i.e. we reward any expected reduction of $f$ in
proportion to the reduction $f_{\rm min} - f(\mathbf{x})$. This can be
evaluated analytically
\begin{align}
  \begin{split}
    \mathcal{A}_{\rm EI}({\mathbf{x}})= {}& \langle {\rm max}(0,f_{\rm min} - f(\mathbf{x})) \rangle = \int_{-\infty}^{\infty} {\rm max}(0,f_{min}-f)\mathcal{N}(f(\mathbf{x});\mu(\mathbf{x}),\sigma(\mathbf{x})^2)\,\, df(\mathbf{x}) = \\
            {}& \int_{-\infty}^{f_{\rm min}} (f_{\rm min} - f) \frac{1}{\sqrt{2\pi \sigma^2}}\exp\left[{-\frac{f-\mu}{2\sigma^2}}\right] \,\,df = \\
            {}& (f_{\rm min} - \mu)\Phi\left(\frac{f_{\rm min} - \mu}{\sigma}\right) + \sigma \phi\left(\frac{f_{\rm min} - \mu}{\sigma}\right) = \sigma \left[ z \Phi(z) + \phi(z) \right],
  \end{split}  
\end{align}
where we dropped the explicit dependence on $\mathbf{x}$ in the third
step, and the cumulative distribution function and standard normal
distribution are denoted $\Phi$ and $\phi$, respectively.  In the last
step we write the result in the standard normal variable
$z=\frac{f_{\rm min}-\mu}{\sigma}$. BayesOpt will exploit regions of
expected improvement when the term $z \Phi(z)$ dominates, while new,
unknown regions will be explored when the second term $\phi(z)$
dominates. For the expected improvement acquisition function, the
exploration-exploitation balance is entirely determined by the set of
observed data $\mathcal{D}_{1:n}$ and the $\mathcal{GP}$ kernel.

The lower confidence-bound acquisition function introduces an additional
parameter $\beta$ that explicitly sets the level of exploration
\begin{align}
  \mathcal{A}(\mathbf{x})_{\rm LCB} = \beta \sigma(\mathbf{x}) - \mu(\mathbf{x}).
\end{align}
The maximum of this acquisition function will occur for the maximum of
the $\beta$-enlarged confidence envelope of the $\mathcal{GP}$. We
use $\beta=2$, which is a very common setting. Larger values of
$\beta$ leads to even more explorative BayesOpt algorithms. 

\begin{figure}
  \begin{center}
    \includegraphics[width=0.46\linewidth]{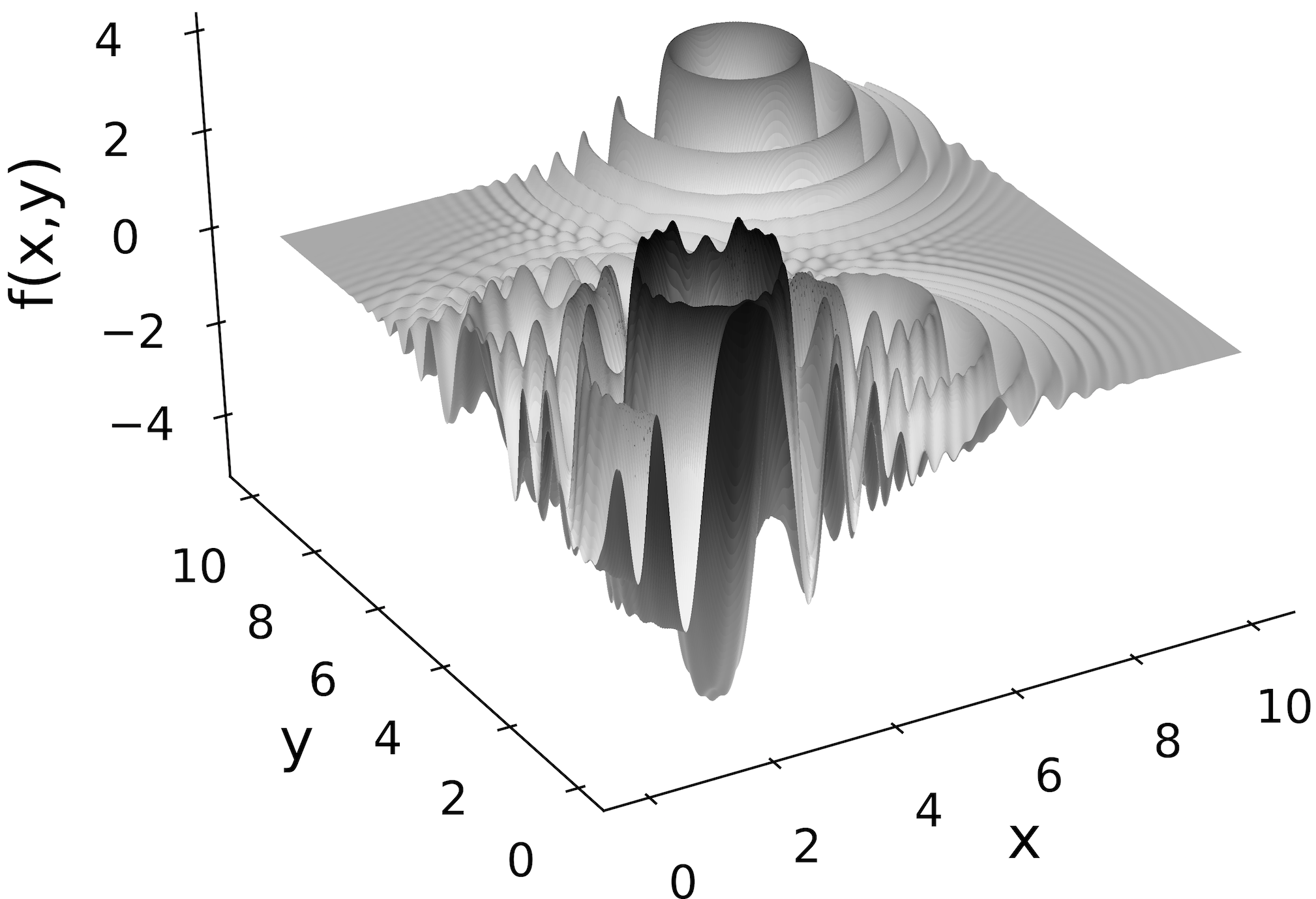}
    \includegraphics[width=0.49\linewidth]{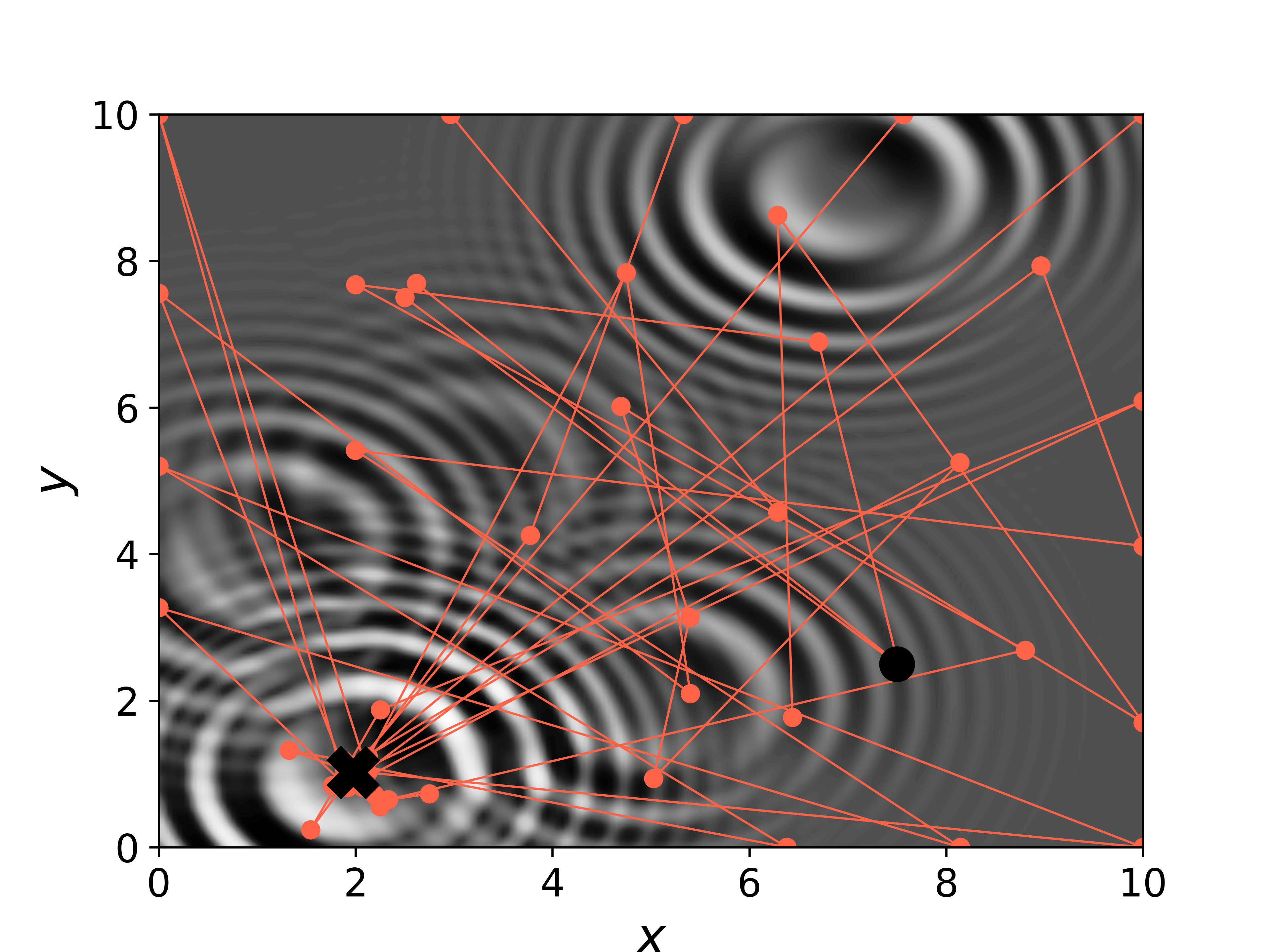}
    \caption{The trace (red dots connected by red lines) of 50
      BayesOpt iterations with a Matern 3/2 $\mathcal{GP}$ kernel and expected improvement
      acquisition function for finding the unique global minimum
      $x_{\star} = (2.002992,1.006096)$ (cross) of the Langermann
      function. The search is initiated at the coordinates
      $[(x_0,y_0);(x_1,y_1)] = [(7.5,2.5);(2.5,7.5)]$. The explorative
      nature of BayesOpt is clearly manifested by the coverage of the
      function evaluations. For this particular setting, the BayesOpt
      algorithm arrives in the vicinity of the optimum already after
      20 iterations.}
    \label{fig:bayesopt_Langermann}
  \end{center}
\end{figure}

Besides being derivative-free---although derivatives can in fact be
incorporated~\cite{NIPS2017_7111}---it is in many ways the explorative
nature of BayesOpt that is most attractive. This feature is most
easily observed in the optimization of a rather complex
two-dimensional function with several local minima, such as the
Langermann function
\begin{align}
  \begin{split}
    f(x,y) =  {}& -\sum_{i=1}^{5}\left\{ \frac{ c_i \cos\left[ \pi((x-a_i)^2 + (y-b_i)^2)\right]  }{  \exp\left[ \frac{1}{\pi} ( (x-a_i)^2 + (y-b_i)^2) \right]  } \right\}, \\
    {}& \mathbf{a} = (3,5,2,1,7), \\
    {}& \mathbf{b} = (5,2,1,4,9),\\
    {}& \mathbf{c} = (1,2,5,2,3).
  \end{split}
\end{align}
In Fig.~\ref{fig:bayesopt_Langermann} we show the sequence
${(x_i,y_i)}_{i=0}^{51}$ of 50 evaluation points following two initial
points of BayesOpt with an Matern 3/2 kernel, expected improvement acquisition function 
without ARD.
It should be made clear that, given a limited computational budget,
the success of BayesOpt hinges on the choice of kernel and acquisition
function. In the example above with a Langermann function, the
non-smooth nature of the Matern 3/2 kernel is advantageous compared
to, e.g., the squared exponential kernel. The exploration-exploitation
balance also leverages the success ratio of BayesOpt. To learn more
about BayesOpt, it is obviously instructive to benchmark and compare
different BayesOpt algorithms using well-known test functions. For
this purpose we first need to select a performance measure for
optimization algorithms.

\section{Measuring the performance of optimization
  algorithms \label{sec:performance}}
To analyze the performance of derivative-free optimization algorithms
we follow Ref.~\cite{wild2009} and define a data profile
\begin{equation}
  d_{s}(\alpha) = \frac{1}{|\mathcal{P}|} {\rm size}\left\{ p \in \mathcal{P}: \frac{t_{s,p}}{D_{p}+1} \leq \alpha \right\}.
\end{equation}
The data profile enables direct comparison between a set of
optimization algorithms $\mathcal{S}$, all of which are applied to a
set of well-defined optimization problems $\mathcal{P}$. For each
$(s,p) \in \mathcal{S} \times \mathcal{P}$, the performance measure
$t_{s,p} > 0$ denotes the number of function evaluations that are
required for optimization algorithm $s$ applied to a problem $p$ to
satisfy some convergence criterion. Thus, $d_{s}(\alpha)$ is the
fraction of problems that can be solved within $\alpha$ function
calls. The performance measure can be further normalized to $D_{p}+1$,
where $D_p$ denotes the dimensionality of the parameter domain in
problem $p$. This is an attempt to account for some of the complexity
due to a larger number of parameters in the problem. This dimensional
scaling is only approximate and motivated by the $(D_p+1)$ function
evaluations required to compute a $D$-dimensional simplex, i.e. a
$D$-dimensional triangle of function evaluations.

Each combination of starting point and objective function
(plus other possible constraints) constitutes a
separate problem $p$. The data profile is a monotonically increasing
function between zero and one, and a large value of $d_{s}(\alpha)$
for small values of $\alpha$ is better. In line with
Ref.~\cite{wild2009} we employ a convergence criterion
\begin{equation}
  f(x_0) -f(x) \geq (1-\tau)(f(x_0)-f_L).
  \label{eq:crit}
\end{equation}
This is fulfilled for any $x$ where the initial function value
$f(x_0)$ is reduced $1-\tau$ times the best possible reduction $f(x_0)
- f_L$. We will set $f_L$ to be the lowest value of $f$ achieved by
any solver $s\in\mathcal{S}$. Although we will come close to the true
solution $x_{\star}$ in a few cases, it is highly unlikely that any
derivative-free solver will arrive at $f(x_{\star})$. For that, one
would typically have to resort to gradient-based optimization
algorithms. With BayesOpt, we will consider a 90\% reduction in the
function value as a sign of convergence, i.e. set $\tau=0.1$. For the
purpose of comparing BayesOpt with other derivative-free optimization
algorithms we will occasionally use $\tau=0.01$. When the objective
functions are represented by known test functions (described in
Sec.~\ref{sec:test_functions} and~\ref{app:testfunctions}), we will
set $f_L = f(x_{\star})$. Throughout this study we will only allow a
maximum of $\alpha=$250 function calls per solver and problem, and set
$t_{p,s} = \infty $ in case the convergence criterion is not
fulfilled. This upper limit on $\alpha$ should cover most scenarios
that would call for the use of BayesOpt.

\subsection{Selecting parameter starting points in $\mathbb{R}^D$}
A multi-dimensional parameter domain $\mathcal{X} \subset
\mathbb{R}^D$ of the objective function leads to the increasingly
likely existence of multiple points corresponding to local and/or
global minimizers. In reality, the choice of initial point, i.e.\ where we
start the optimization run, will determine the local minimum that the
optimizer converges to. When benchmarking, it will be necessary to
start each solver $s$ from $N>1$ starting points in $\mathcal{X}$. In
this work, we use an identical set of $N=2D$ starting points for all
optimizers in a $D-$dimensional domain. This is somewhat motivated by
Wendel's theorem~\cite{wendel} that states that for $N$ random points
on the $D-$sphere, the probability $P(N,D)$ that they all lie on some
hemisphere is given by
\begin{equation}
  P(N,D) = 2^{-N+1}\sum_{k=0}^{D-1} \left( \begin{array}{c}
    N-1\\
    k
  \end{array}
    \right).
\end{equation}
Thus for random $N=2D$ points on the $D-$sphere,
$P(2D,D)=0.5$. Although we do not distribute points on a hypersphere,
we use it to motivate a rule of thumb. It should be pointed out that
some authors argue for a slightly more expensive rule of thumb to
select $N=10D$ starting points when initializing computer simulations
in $\mathbb{R}^D$~\cite{Loeppky}.

\begin{figure}
  \begin{center}
    \includegraphics[width=\textwidth]{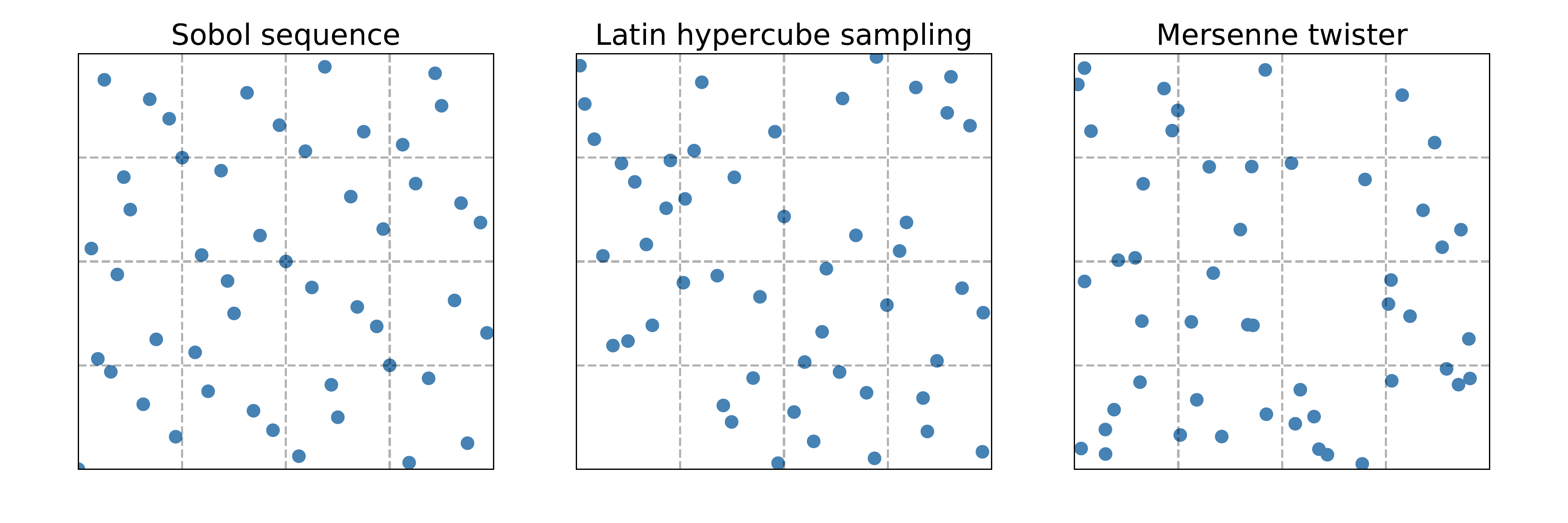}
    \caption{50 points in $\mathbb{R}^2$ according to the Sobol
      low-discrepancy sequence, latin hypercube sampling (LHS), and
      the standard Mersenne twister pseudo-random generator. The Sobol
      sequence minimizes the occurrence of large gaps
      (discrepancy). In fact, given the grid decomposition of the
      area, LHS produces one empty region in the lower left corner,
      and the Mersenne twister produces two empty regions.}
    \label{fig:randomness}
  \end{center}
\end{figure}

A priori, we do not distinguish between different parts of the
parameter domain, so we select the $N$ starting points using a
space-filling algorithm in the form of a quasi-random Sobol
sequence. We employ a Python implementation of this algorithm. The
mathematical underpinnings of the Sobol sequence are provided in the
original paper~\cite{sobol} and a discussion related to the numerical
implementation is given in e.g.\ Ref.~\cite{Bratley}. This is a
so-called low-discrepancy sequence, which in fact is the opposite of a
random sequence. It is designed to generate each successive sample
point as far away as possible from all existing sample points. This
tends to sample the space more uniformly than pseudo-random numbers,
at least for lower-dimensional domains. We will not go beyond $D=12$
in any part of this work. Although the Sobol sequence can exhibit gaps
in multi-dimensional spaces, it has several advantages. In addition to
fast generation, a Sobol sequence is straightforward to augment with
additional sample points. We remind the reader that Latin hypercube
sampling (LHS)~\cite{mckay}, which is a different kind of algorithm
for generating space-filling samples, is in most cases not easy to
augment while preserving the latin hypercube structure. The
space-filling differences between Sobol, LHS, and conventional
pseudo-random numbers in $\mathbb{R}^2$ are illustrated in
Fig.~\ref{fig:randomness}.

\section{Analyzing a set of test functions \label{sec:test_functions}}
\begin{figure}
  \begin{center}
    \includegraphics[width=0.43\linewidth]{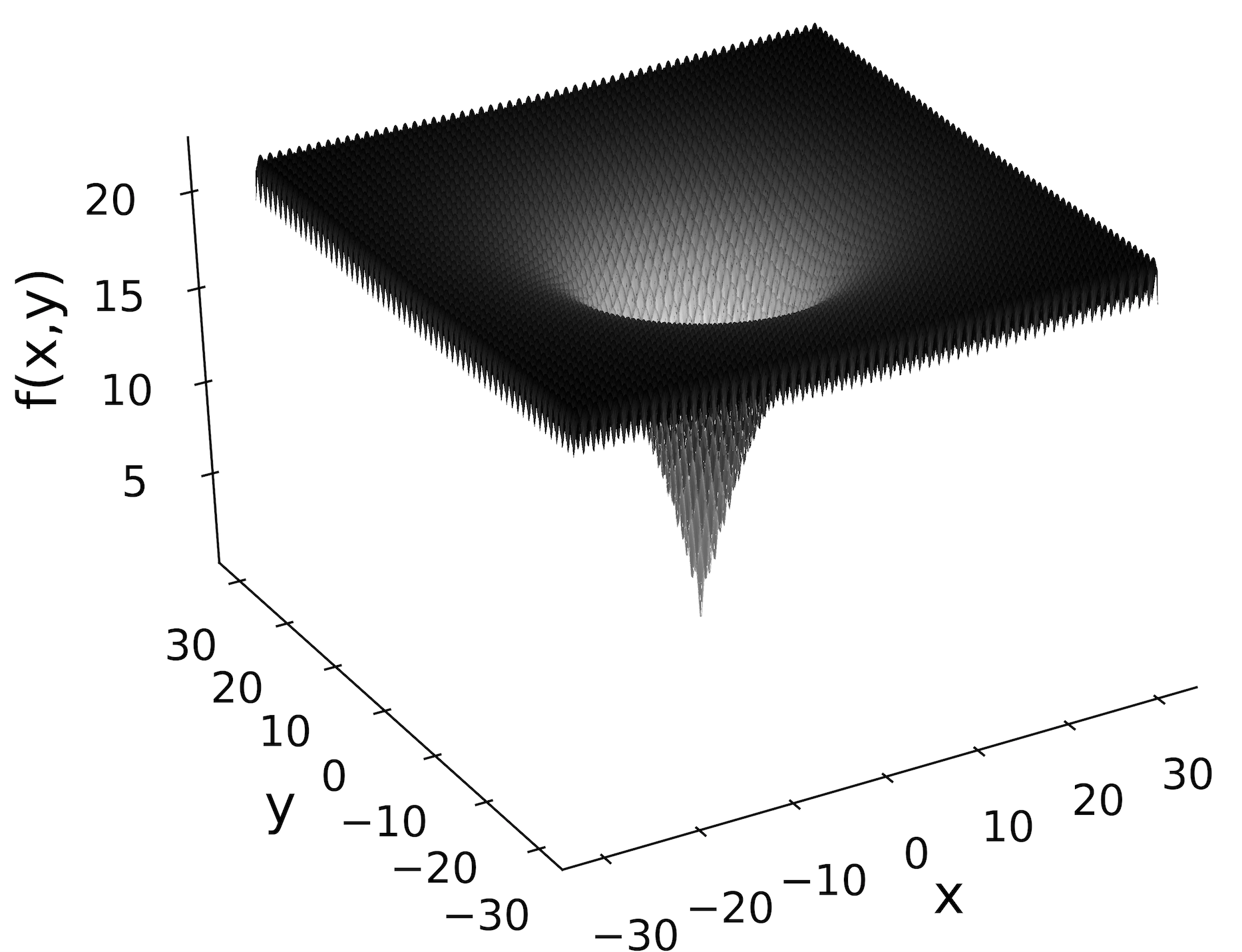}
    \includegraphics[width=0.46\linewidth]{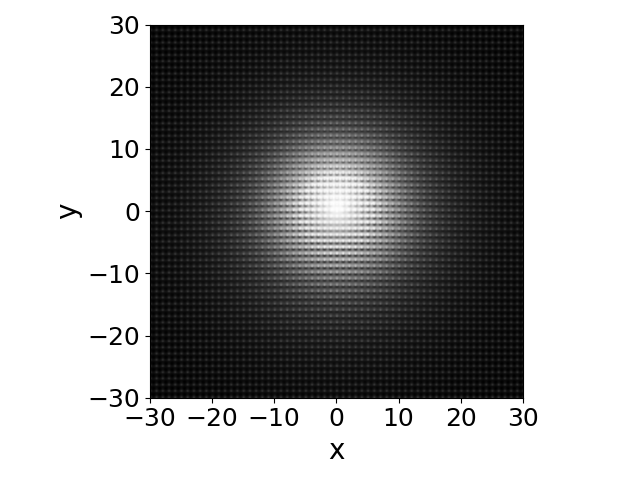}
    \includegraphics[width=0.43\linewidth]{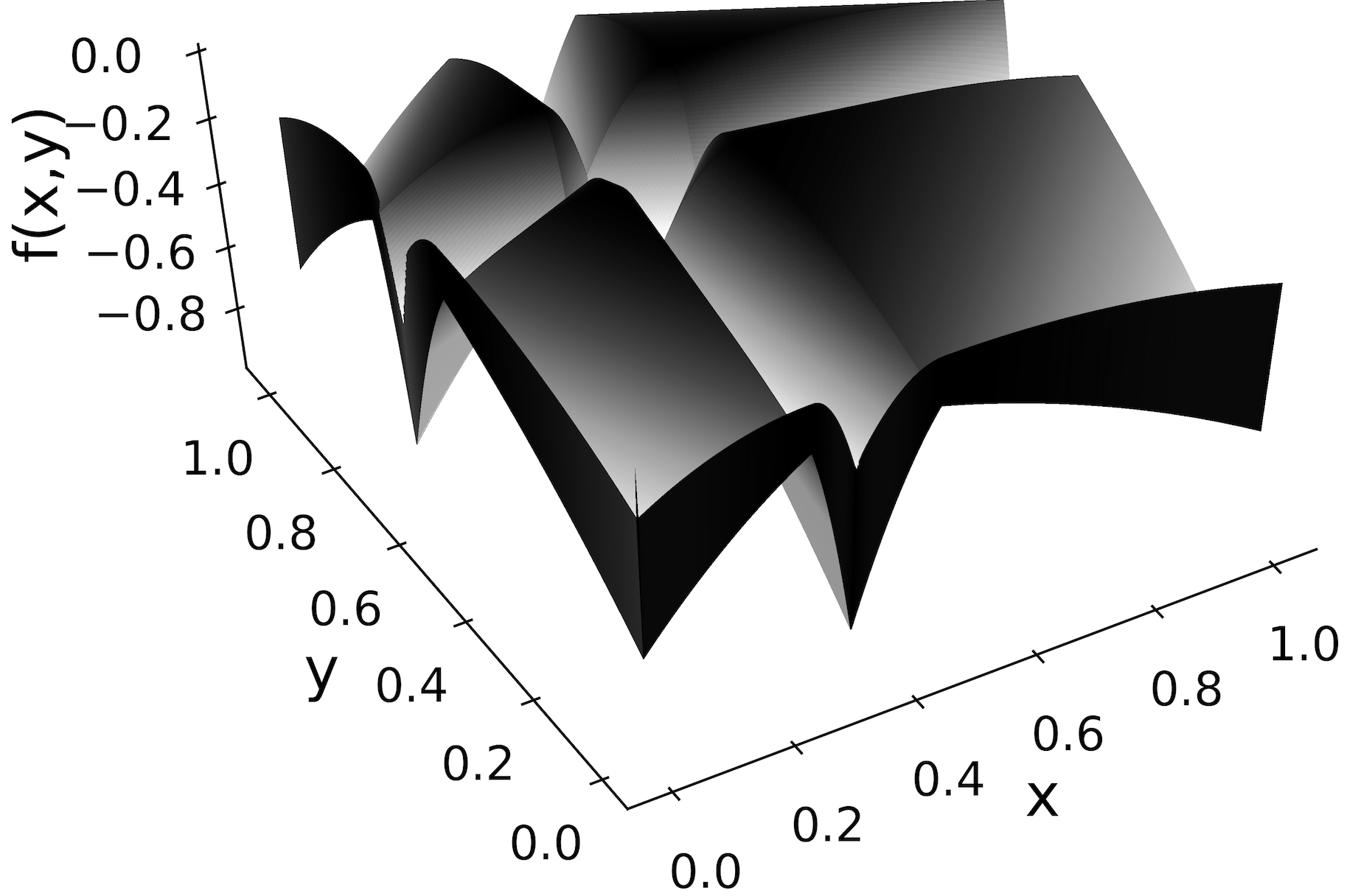}
    \includegraphics[width=0.46\linewidth]{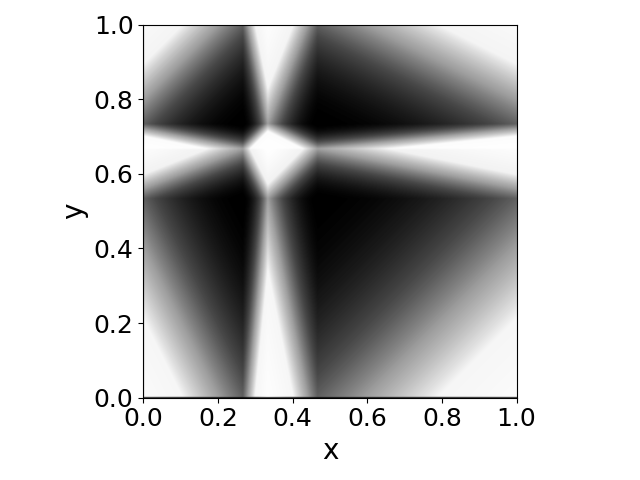}
    \includegraphics[width=0.43\linewidth]{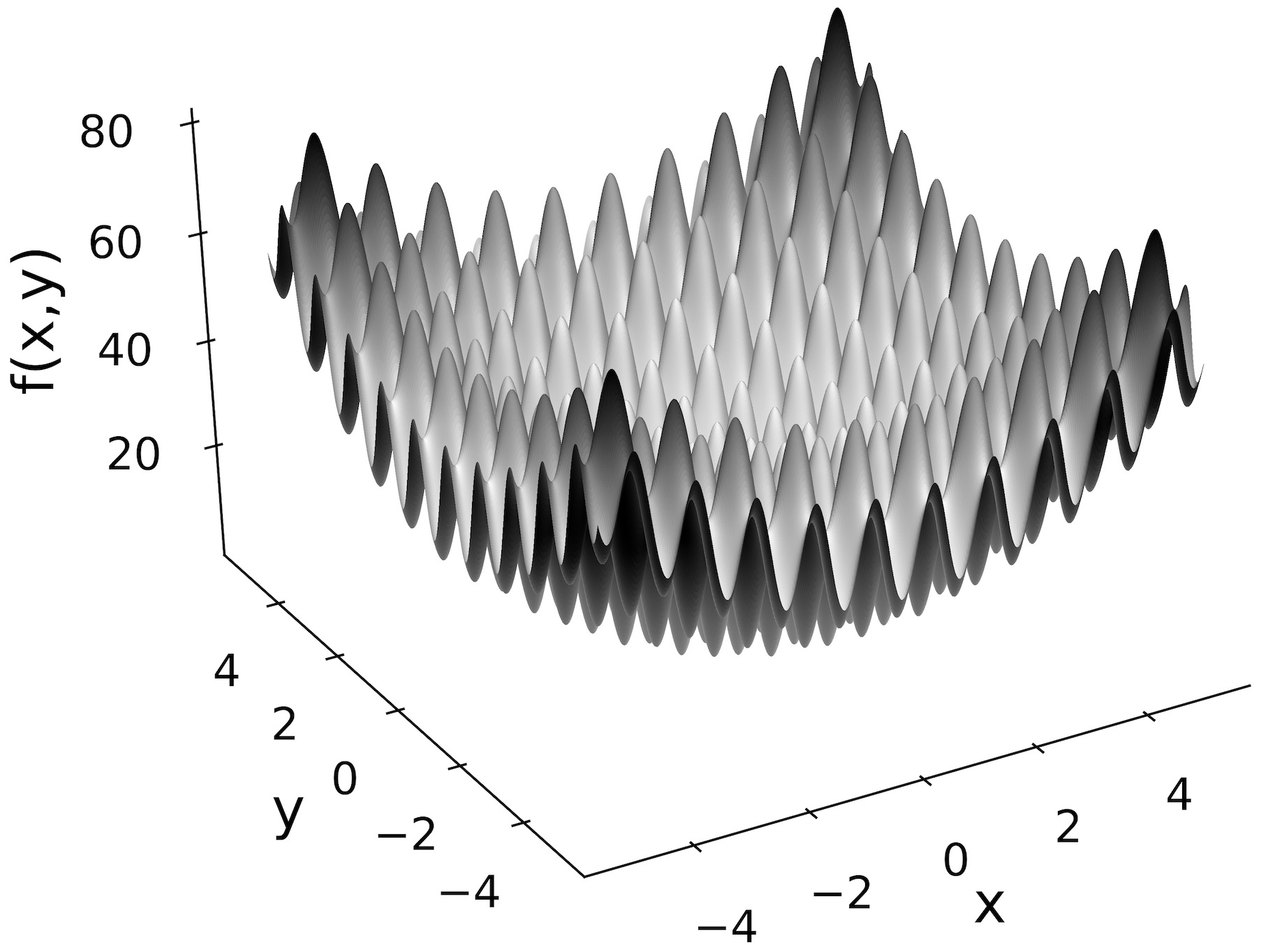}
    \includegraphics[width=0.46\linewidth]{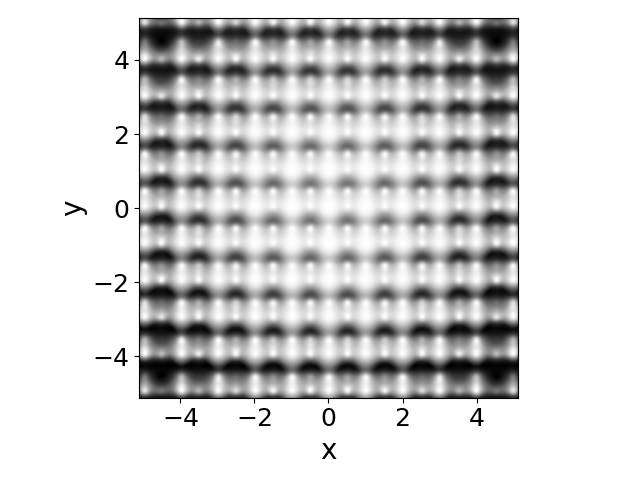}
    \caption{Surface plots and projections for the test functions with
      $D=2$. One
      test function per row, top to bottom: Ackley, Deceptive, and
      Rastrigin. See the text and \ref{app:testfunctions} for details.}
    \label{fig:test_functions_A}
  \end{center}
\end{figure}

\begin{figure}
  \begin{center}
    \includegraphics[width=0.43\linewidth]{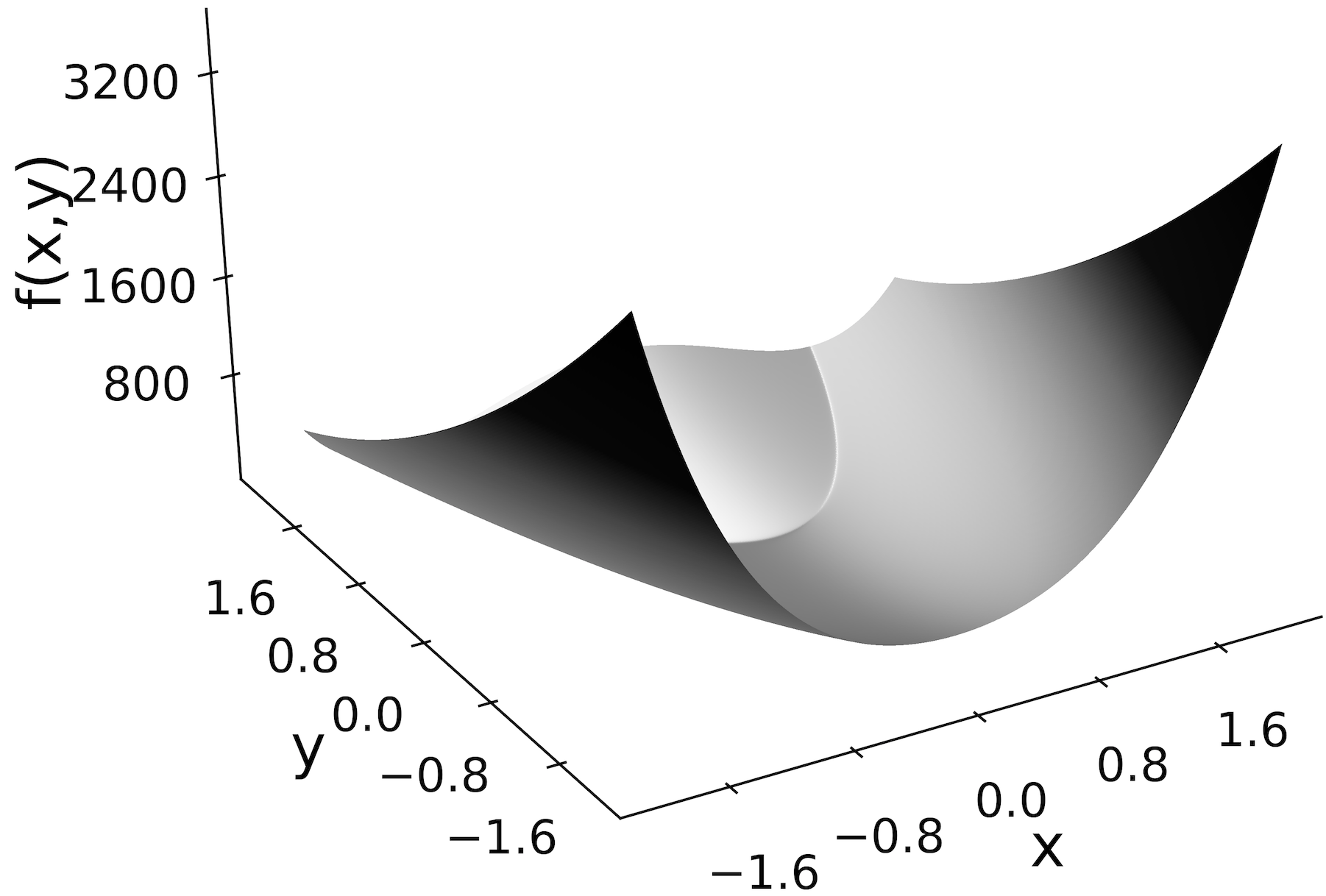}
    \includegraphics[width=0.46\linewidth]{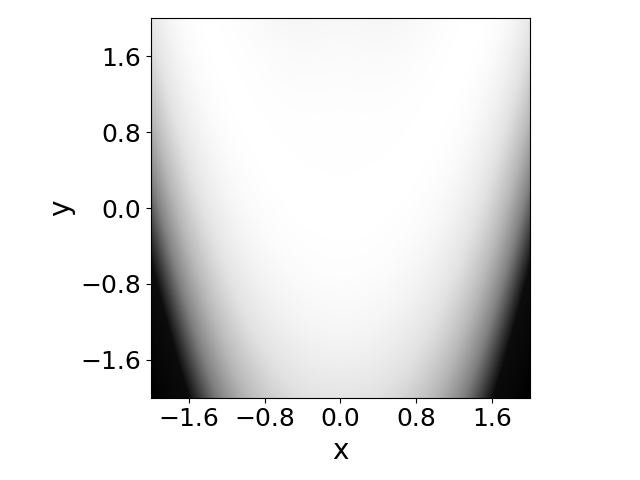}
    \includegraphics[width=0.43\linewidth]{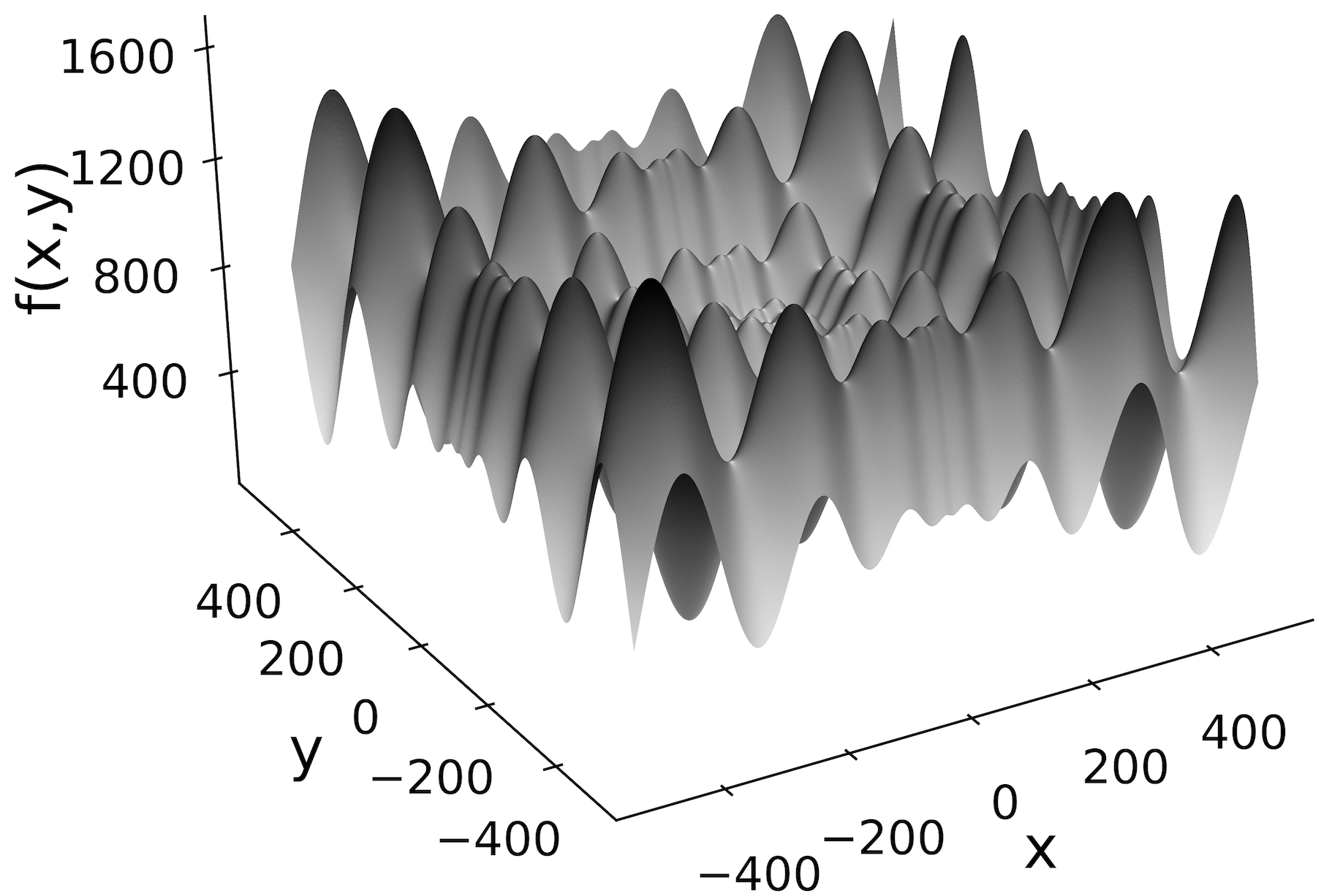}
    \includegraphics[width=0.46\linewidth]{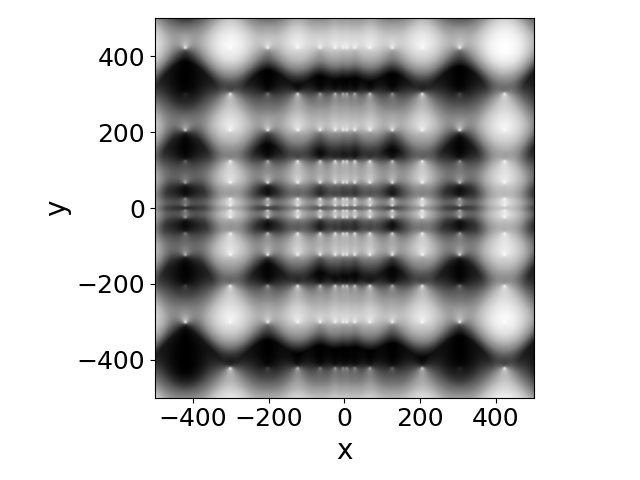}
    \includegraphics[width=0.43\linewidth]{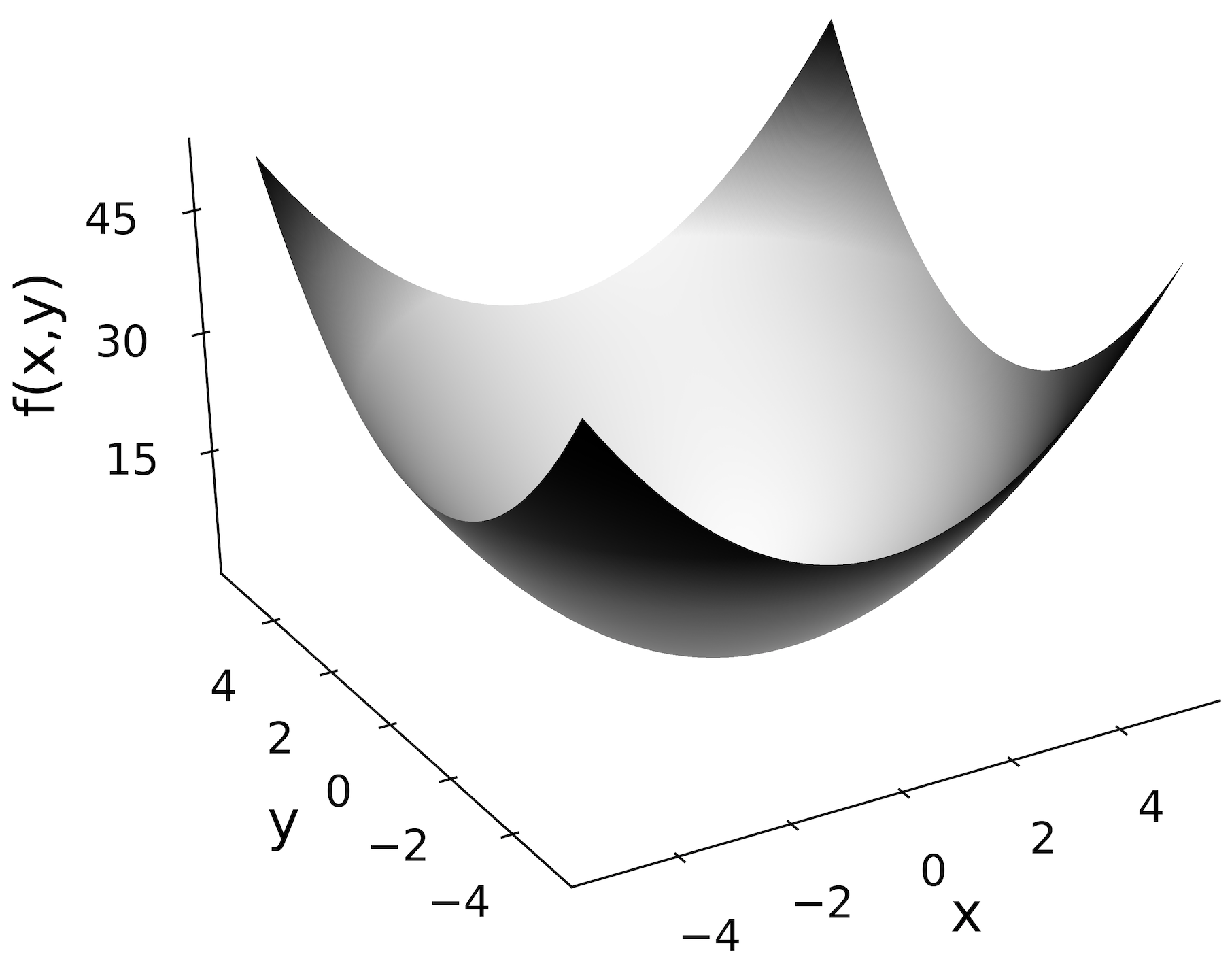}
    \includegraphics[width=0.46\linewidth]{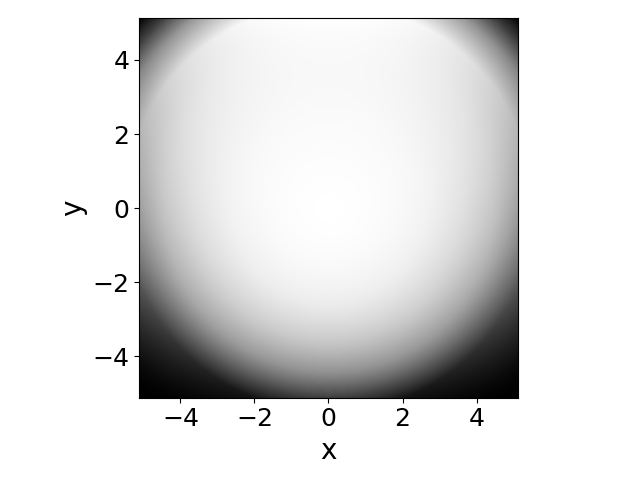}
    \caption{Surface plots and projections for the test functions with
      $D=2$. One
      test function per row, top to bottom: Rosenbrock, Schwefel, and
      Sphere. See the text and \ref{app:testfunctions} for details.}
    \label{fig:test_functions_B}
  \end{center}
\end{figure}

Before we tackle an optimization problem in nuclear physics, we will
explore and benchmark the performance of BayesOpt on a set of test
functions. To this end, we have selected a set of six multivariate and
continuous functions $f: \mathbb{R}^{D} \rightarrow \mathbb{R}$, each
defined on some domain $\mathcal{X} \subset \mathbb{R}^D$. The
functions are defined for any $D>0$, but we will only consider
$D=2,4,8$. The set of test functions reflects an average of various
spatial characteristics. Two-dimensional graphical representations are
shown in Figs.~\ref{fig:test_functions_A} and \ref{fig:test_functions_B},
with explicit expressions given in~\ref{app:testfunctions}.

A comparison of two or more optimization algorithms on a finite set of
test functions is neither fair nor conclusive. Indeed, although one
algorithm (or class of methods) appears to be more successful in
finding optima, it is clear from the \textit{no free lunch} theorem in
optimization that the averaged performance of any two algorithms on
the set of all possible functions will be the same. Here, we merely
set out to compare and analyze the performance of BayesOpt on a
limited set of characteristically different continuous test functions.

\begin{figure}
  \begin{center}
    \includegraphics[width=0.95\textwidth]{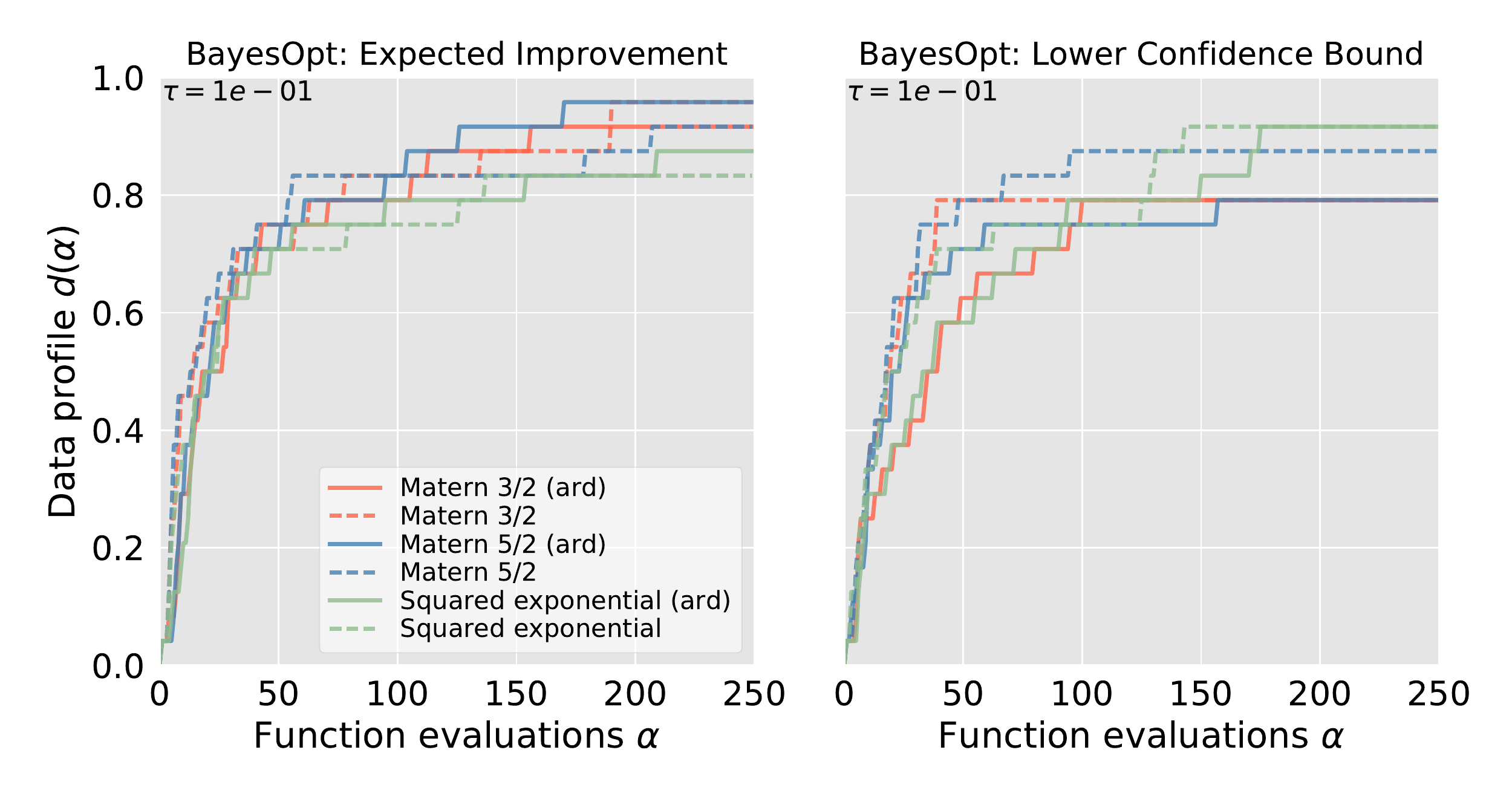}
    \includegraphics[width=0.95\textwidth]{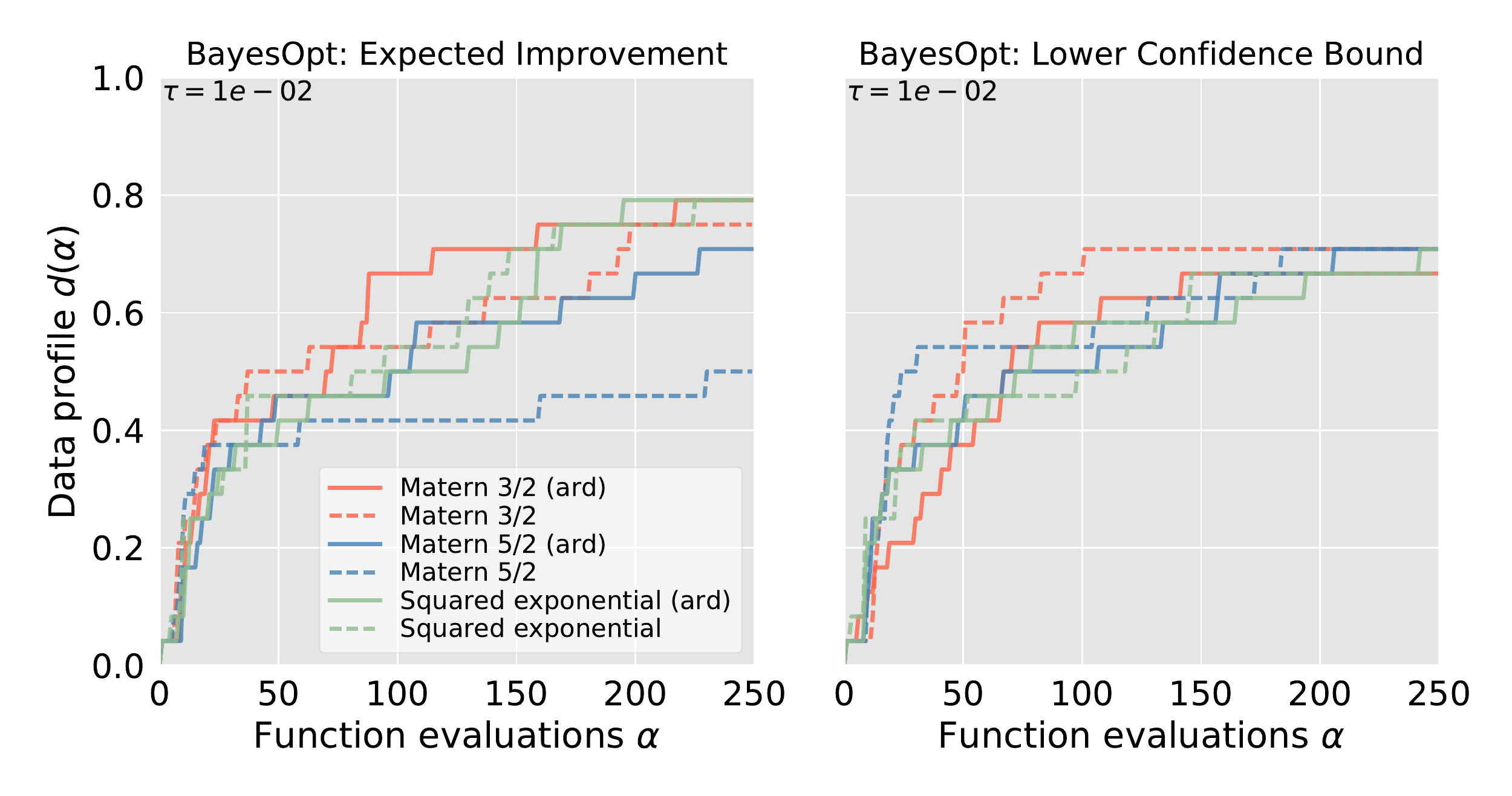}
    \caption{Data profiles (measured with our six test functions in $D=2$
      domains) for BayesOpt using three different kernels (see legend), with and without
      ARD, and using two different acquisition functions: 
      expected improvement (left column), lower confidence-bound (right column). The convergence criterion, set by $\tau$,
      corresponds to a $(1-\tau)\cdot 100$\% reduction of the initial
      function value. Data profiles for
      $\tau=0.1$ (top row). Data profiles for $\tau=0.01$ (bottom row).}
    \label{fig:prof_test_2d}
  \end{center}
\end{figure}

We now turn our attention to the resulting data profiles when applying
BayesOpt for finding the minimizer in each one of these test functions
with parameter domains in $D=2,4,8$ dimensions. In total, we will
analyze 12 BayesOpt algorithms composed from combining three kernels,
two acquisition functions, and with or without ARD. We use a Sobol
sequence to initiate each BayesOpt algorithm at $N=2D$ different
starting points in each parameter domain in $\mathbb{R}^D$. Remember
that we refer to each specific combination of starting point and test
function as a \textit{problem}. Combined, with the different solvers
(optimization algorithm with specific settings) this is a rather large
dataset and we analyze it from several different angles. The data
profiles for different versions of BayesOpt applied to all six test
functions in $D=2$, see Fig.~\ref{fig:prof_test_2d} (top row), indicate
that $d(50)\sim 0.7$, i.e. that $\sim 70$\% of the problems are
converged at the $\tau=0.1$ level within 50 function evaluations. The
performances of the expected improvement and the lower
confidence-bound acquisition functions are very similar. Although,
expected improvement exhibits a slightly better and more coherent
performance across all $\mathcal{GP}$ kernels, and at 150 function
evaluation more than 80\% of the test functions in $D=2$ are
converged.
ARD, i.e. to learn each individual length scale hyperparameter in the
$\mathcal{GP}$ kernel, is often an efficient way of pruning irrelevant
features. We find that exploiting ARD increases the performance of
BayesOpt once data from a sufficient number of function evaluations has informed
the $\mathcal{GP}$ kernel on the possible anisotropic structure of the
objective function. This becomes even clearer if we enforce a
stronger convergence criterion with $\tau=0.01$, see
Fig.~\ref{fig:prof_test_2d} (bottom row). There seems to be a slight
advantage of using ARD with the expected improvement acquisition function.

\begin{figure}
  \begin{center}
    \includegraphics[width=0.95\textwidth]{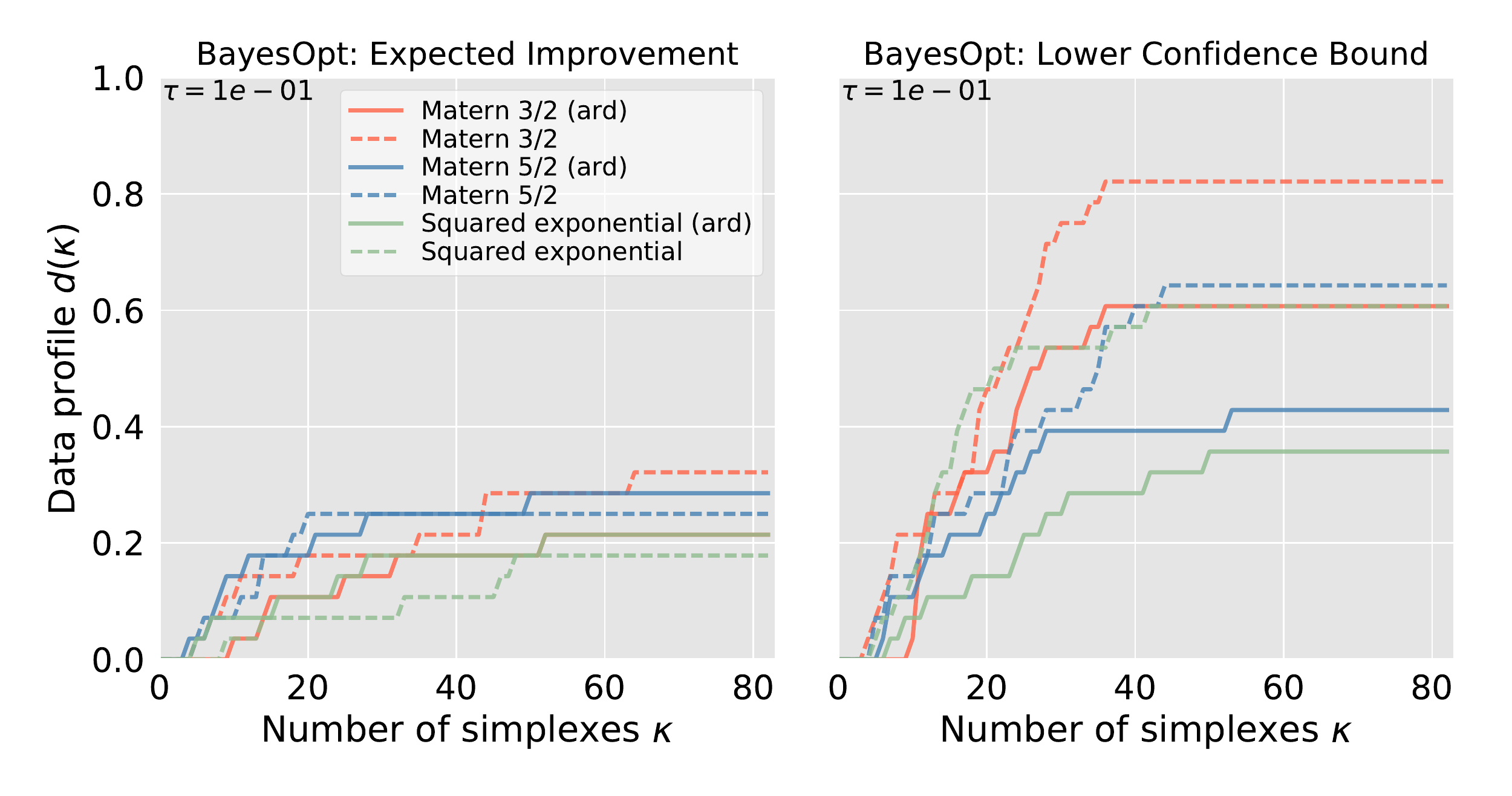}
    \includegraphics[width=0.95\textwidth]{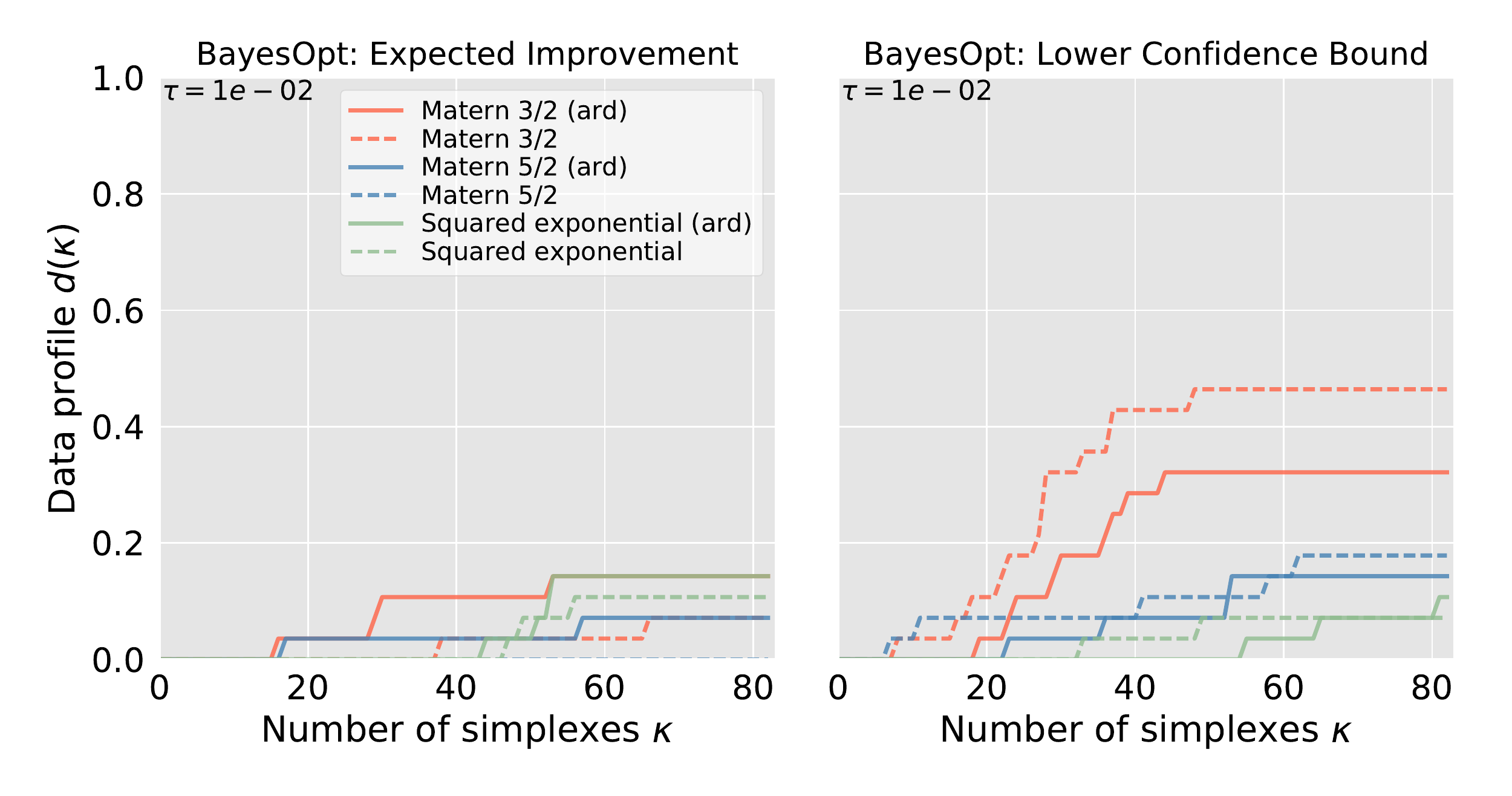}
    \caption{Data profiles for the Ackley test
      function.  The
      performances are averaged over parameter domains with
      $D=2,4,8$, and we employ a dimensionality normalization
      $\kappa = \alpha / (D_p+1)$.}
    \label{fig:prof_Ackley}
  \end{center}
\end{figure}

In $D=2$, and for this set of functions, the Matern kernels perform
slightly better than the squared exponential. In general, the Matern
kernels are better tailored to non-smooth objectives. This becomes
even clearer when we study the performance of all BayesOpt algorithms
on the Ackley function (one hole on a semi-flat surface with several
periodic shallow local minima) in $\mathbb{R}^{D=2,4,8}$, see
Fig.~\ref{fig:prof_Ackley}.
From the characteristics of the data profiles shown in
Fig.~\ref{fig:prof_Ackley}, it is obvious that the exploratory nature
of the lower confidence-bound acquisition function is highly
advantageous for finding the global minimum hiding on a surface
covered with local minima. Clearly, it is important to tailor the BayesOpt
acquisition function and underlying $\mathcal{GP}$ kernels to the
spatial structure of the objective function. This result also reflects
the two fundamental and competing aspects of BayesOpt. If we
incorporate prior beliefs, or even knowledge, about the objective
function, then BayesOpt will perform rather well. On the other hand,
the usefulness of BayesOpt will consequently be limited by the
arbitrariness and uncertainty of \textit{a priori} information. This
is further complicated by the fact that we typically resort to
BayesOpt when we know very little about the objective function in the
first place, since it is computationally expensive to evaluate.

\begin{figure}
  \begin{center}
    \includegraphics[width=0.95\textwidth]{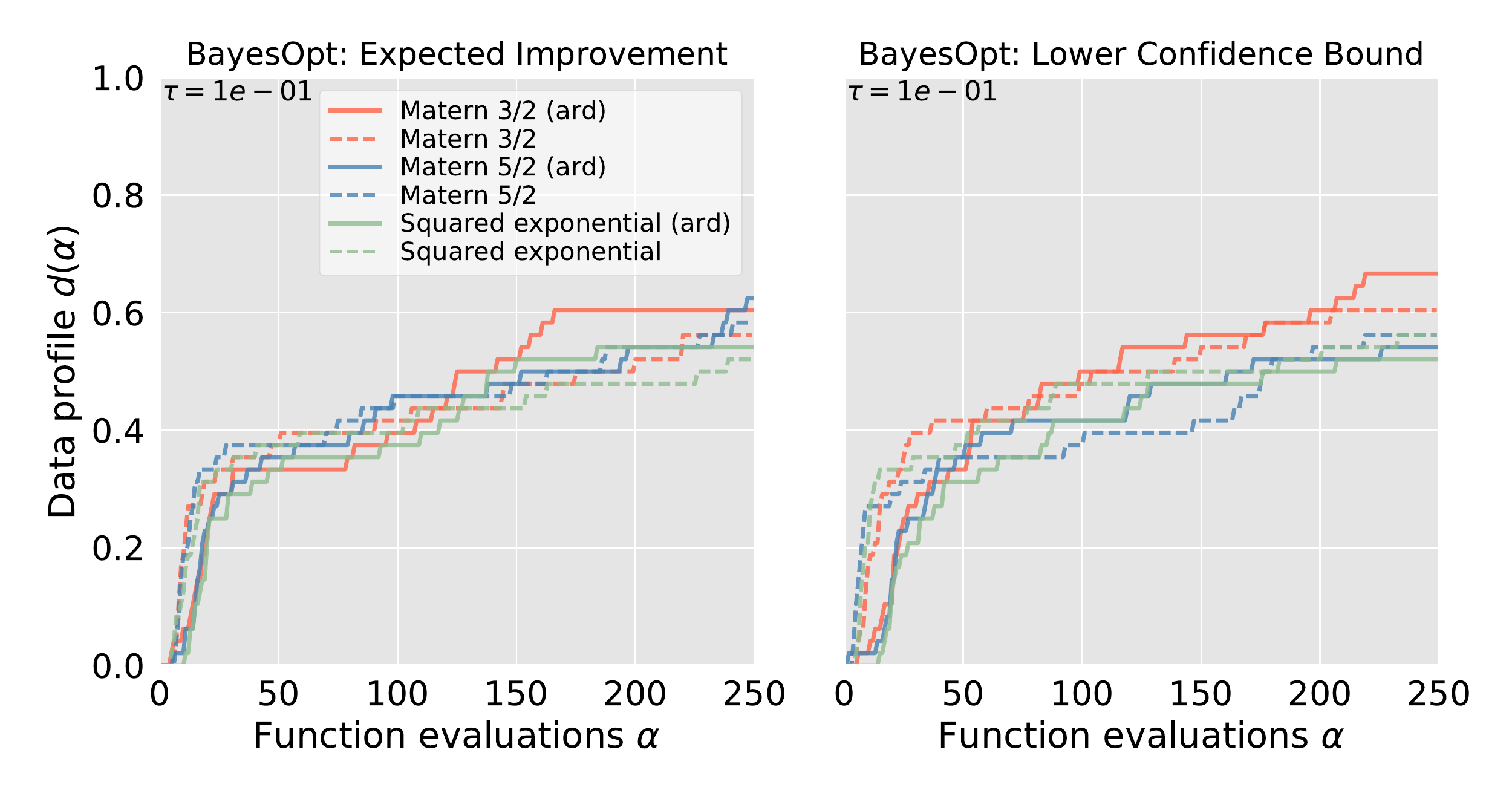}
    \includegraphics[width=0.95\textwidth]{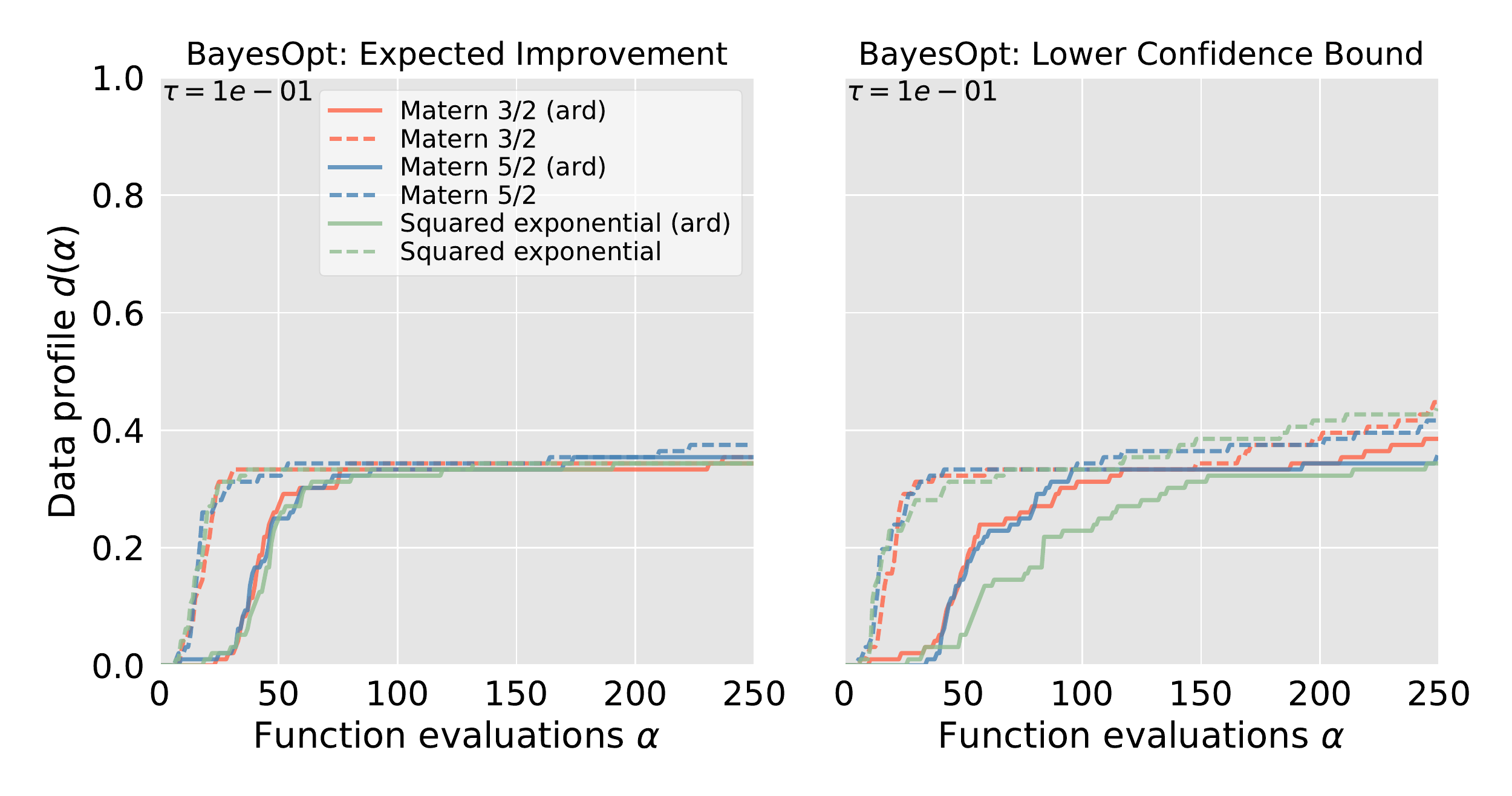}  
    \caption{Data profiles for BayesOpt applied to all test
      functions in $D=4$ dimensional domains (top row) and $D=8$
      dimensional domains (bottom row).}
    \label{fig:4d_8d}
  \end{center}
\end{figure}

Data profiles for the test functions as we increase the dimensionality
of the parameter domain from $D=2$ (in Fig.~\ref{fig:prof_test_2d}) to $D=4$ and $D=8$ are shown in
Fig.~\ref{fig:4d_8d}.
We can conclude that having a larger number of parameters provides a
significant challenge to BayesOpt, as for all optimization
algorithms. Indeed, in $D=4$ dimensions it takes more than
$\alpha=150$ function evaluations to reach a data profile value of $d
\sim 0.5$. The more exploratory lower confidence-bound acquisition
function exhibits a slightly larger performance spread with respect to
different $\mathcal{GP}$ kernels. This becomes even more prominent as
we increase the dimensionality of the parameter domain to $D=8$. For
the set of test functions that we have employed it is marginally
advantageous to be more exploratory for higher dimensional objective
functions. We also note that the potential benefit of ARD requires
data from more than $\alpha=250$ function evaluations. This is natural
since ARD introduces more hyperparameters that need to be
determined. For all problems in $D=4,8$ dimensions, BayesOpt with
Matern kernels converge faster than BayesOpt with a squared
exponential kernel. As above, the squared exponential kernel does not
capture the high-frequency modes that are present in some of the
functions.

\section{Bayesian optimization of the nucleon-nucleon interaction\label{sec:NP}}

An EFT of the nuclear interaction essentially corresponds to a
low-energy parameterization of QCD in a fashion that is consistent with
the symmetries of the more fundamental theory. To devise a true EFT
description of the nuclear interaction is a major intellectual
challenge in \textit{ab initio} nuclear theory. Several avenues are
currently being explored. The physical underpinnings of the objective
function we seek to minimize are described in somewhat more detail
in~\ref{app:nnlo}. Regarding the optimization of the
LECs, the inclusion of calibration data from more complex atomic
nuclei and low-energy nuclear processes, e.g. $NNN$ scattering, in the
objective function are currently being considered in the
community. Such extensions of the calibration data clearly increases
the information content---but does so at the expense of an increased
complexity in the computer modeling.

A specific aim of this work is to analyze the performance of BayesOpt
for determining the parameters $\mathbf{x}$ in an EFT model of the
$NN$ interaction. For this, we use the proton-neutron sector of an EFT
at NNLO with 12 parameters and try to find the vector of model parameters
$\mathbf{x}$ that are in agreement with existing experimental data on
proton-neutron scattering cross sections. This type of observable is a
measure of the probability of an incident particle (a proton) with a
given kinetic energy to scatter into a certain solid angle element due
to mutual interaction with the target particle (a
neutron\footnote{Since free neutrons decay within $\sim 10$ minutes,
  the neutron target is a composite material containing
  neutrons.}). We have deliberately chosen this class of calibration
data since it does not render particularly challenging model
evaluations. One evaluation of the full objective function
$f(\mathbf{x})$  takes a merely couple of seconds on a standard
desktop. Still, the complexity of the physical model provides a
non-trivial testing ground for assessing nascent applications of
BayesOpt in \textit{ab initio} nuclear physics.

The experimental dataset is composed of $N_G$ groups of measurements
where the $g$-th group consists of $N_{g,d}$ measured cross sections,
with associated random measurement uncertainty,
$\mathcal{O}_{g,i}^{\rm experiment} \pm \sigma_{g,i}$, for $i=1,
\ldots, d$, with a common normalization constant $\nu_g$ and
corresponding experimental systematic error $\sigma_{g,0}$. We employ
the measurement errors as reported by each experimenter. Each group of
data originates from a specific experiment. We restrict the inclusion
of experimental data in the present case to laboratory scattering
energies below 75 MeV. This ensures a rather fast (seconds) evaluation
of the objective function and therefore enables a more detailed
analysis of BayesOpt. Experimental errors across measurement groups
are considered independent and identically distributed. The
normalization constant, together with its uncertainty, represents the
systematic uncertainty of the measurement. For an absolute measurement,
the normalization is given by $\nu_g=1 \pm 0$. Usually this means that
the statistical and systematic errors have been combined with
$\sigma_{g,d}$. Certain experiments are not normalized at
all. Instead, only the angular- or energy-dependence of the cross
section was determined. For these groups of data, $\nu_g$ is solved
for in the optimization by minimizing the discrepancy between the
model prediction $\mathcal{O}_{g,d}^{\rm model}(\mathbf{x})$ and the
experimental data points $\mathcal{O}_{g,d}^{\rm experiment}$. For
such freely renormalized data, the optimal $\nu_g$ is easily obtained
in closed form. For practical purposes, the normalization error can be
considered infinite in these cases, and will therefore not contribute
to the objective function.

In summary, we seek to find the parameter vector $\mathbf{x}$ that
minimizes the deviation between the model and the experimental data,
as measured by the objective function $f$ defined below:
\begin{equation}
  f(\mathbf{x}) = \sum_{g=1}^{N_G}\underset{\nu_{g}}{\rm min}\left\{ \sum_{i=1}^{N_{g,d}} \left( \frac{\nu_{g}\mathcal{O}^{\rm model}_{g,i}(\mathbf{x}) - \mathcal{O}^{\rm experiment}_{g,i}}{\sigma_{g,i}} \right)^2 + \left( \frac{1-\nu_{g}}{\sigma_{g,0}}\right)^2\right\}.
  \label{eq:nnlo_objective}
\end{equation}

This type of (non-linear) least-squares objective function, where we
assume a normal model for the parameters, appears naturally in most
parameter estimation problems. In a setting defined by $\chi$EFT it
would be natural to also incorporate a theoretical model discrepancy
term motivated by the low-energy EFT expansion
itself~\cite{Carlsson:2016,   Melendez, PhysRevLett.115.122301, Epelbaum2015, Furnstahl}. However, in this paper we will focus on
the challenges of mathematical optimization that are associated with
a numerically costly objective function, and for simplicity therefore
only incorporate the experimental errors of the data.

We define three different parameter domains for minimizing the
objective function; \textit{small}, \textit{medium}, and
\textit{large}, see Tab.~\ref{tab:domains}. They differ by the level
of included prior knowledge regarding the range of plausible parameter
values.
\begin{table}
  \begin{center}
  \begin{tabular}{l|c|c|c}
             & \multicolumn{3}{c}{Domains}\\
            \cline{2-4}
    parameter              & small         & medium        & large         \\ \hline
    $\tilde{C}_{^1S_0}^{(np)}$ & (-0.2,-0.1)   & (-5,+5)   & (-5,+5)   \\
    $\tilde{C}_{^3S_1}$       & (+2,+3)       & (-5,+5)       & (-5,+5)        \\
    $C_{^1S_0}$               & (-0.2,-0.1)   & (-5,+5)   & (-5,+5)    \\
    $C_{^3S_1}$              & (-1,+1)       & (-5,+5)       & (-5,+5)       \\
    $C_{^3P_{0}}$             & (-1,+1)       & (-5,+5)       & (-5,+5)       \\
    $C_{^3P_{1}}$             & (-1,+1)       & (-5,+5)       & (-5,+5)       \\
    $C_{^3P_{2}}$             & (-1,+1)       & (-5,+5)       & (-5,+5)       \\
    $C_{^1P_1}$              & (-1,+1)       & (-5,+5)       & (-5,+5)       \\
    $C_{E_1}$                & (-1,+1)       & (-5,+5)       & (-5,+5)       \\
    $c_1$                   & (-0.76,-0.72) & (-0.76,-0.72) & (-5,+5)  \\
    $c_3$                   & (-3.66,-3.56) & (-3.66,-3.56) & (-5,+5)  \\
    $c_4$                   & (+2.41,+2.47) & (+2.41,+2.47) & (-5,+5)  \\ 
  \end{tabular}
  \caption{Three different parameter domains (small, medium, large) of
    the 12 parameters (LECs) in the nuclear physics objective function
    we study here. The LECs govern the short-range contact potential
    ($\tilde{C}_{\cdot}$) at leading order (unit: 10$^4$~GeV$^{-2}$),
    the short-range contact potential ($C_{\cdot}$) at next-to-leading
    order (unit: 10$^4$~GeV$^{-4}$), and the sub-leading long-ranged
    pion potential ($c_{\cdot}$) at next-to-next-to-leading order
    (unit: GeV$^{-1}$). }
  \label{tab:domains}
  \end{center}
\end{table}
The limits of the \textit{large} parameter domain is based on the most
naive estimate without any significant prior information. In contrast,
the limits of the \textit{medium} domain are partly informed by prior
data. Specifically, the three parameters (LECs $c_1, c_3, c_4$)
associated with the long-range part of the nuclear interaction are
constrained to ranges determined from a separate analysis of
pion-nucleon scattering
data~\cite{Hoferichter:2015ft,Hoferichter:2016bc}. The \textit{small}
domain is further informed by previous experience of typical values
for the LECs in the short-ranged contact potential. 
\begin{figure}
  \begin{center}
    \includegraphics[width=0.95\textwidth]{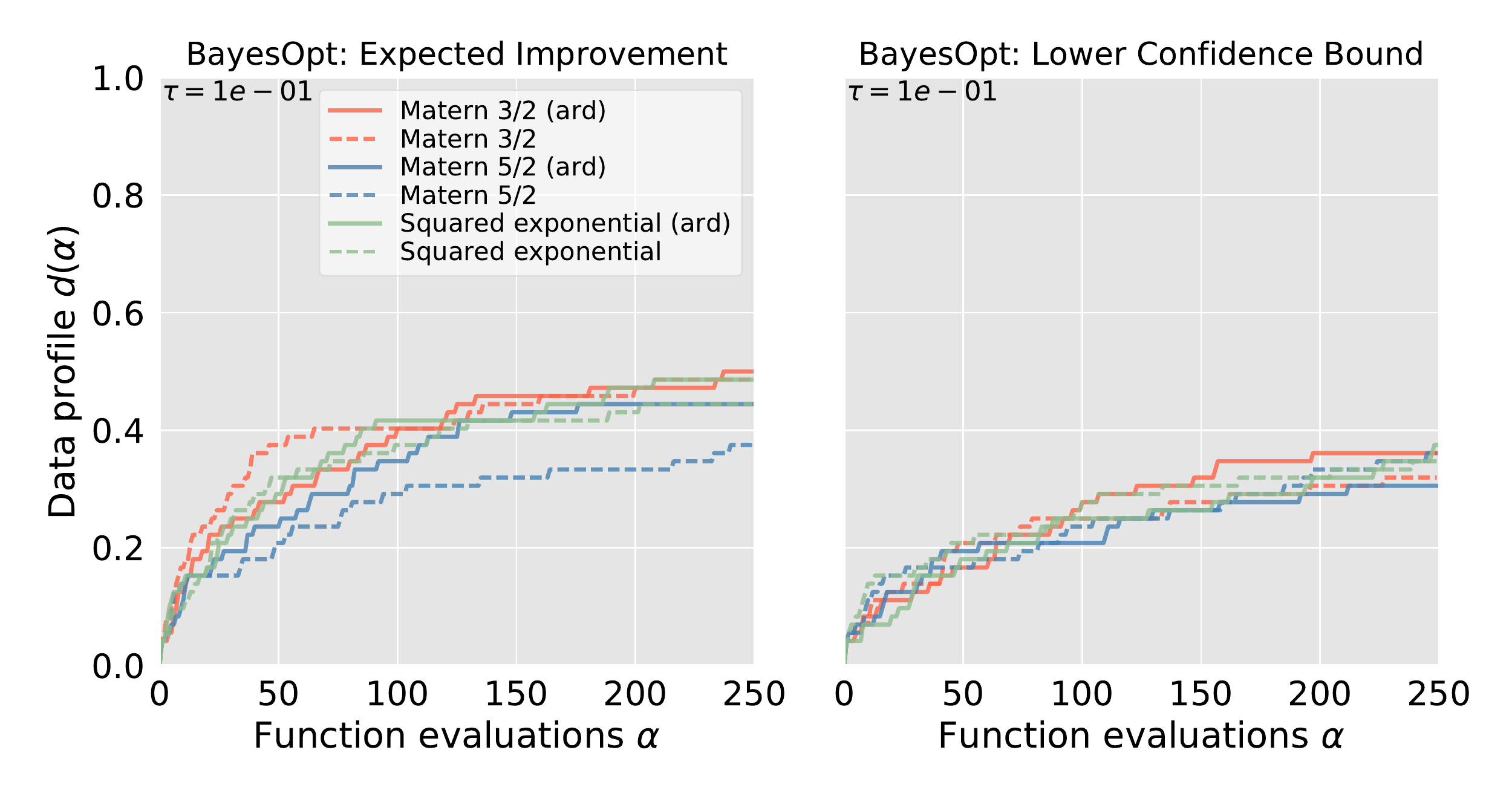}
    \caption{Data profiles for BayesOpt applied to the
      nuclear physics objective function in
      Eq.~\ref{eq:nnlo_objective}, over all parameter domains in
      Tab.~\ref{tab:domains}. The expected improvement acquisition function for all
      kernels, except Matern 5/2, exhibits a better performance compared to
      the lower confidence-bound acquisition function. Improving the performance for lower confidence-bound
      will require further function evaluations.}
    \label{fig:prof_nnlo}
  \end{center}
\end{figure}

The data profiles of the BayesOpt algorithms applied to all domains of
the nuclear physics objective function in Eq.~\ref{eq:nnlo_objective}
are plotted in Fig.~\ref{fig:prof_nnlo}. Clearly, the expected
improvement acquisition function performs slightly better than the
lower confidence-bound acquisition function. It is difficult to draw
any conclusions regarding an optimal choice of kernel. Compared to the
test functions, the squared exponential kernel seems to work a little
bit better than the Matern 5/2 kernel. This result is also somewhat
surprising since the performance of the squared exponential kernel is
nearly identical to the Matern 3/2 kernel.
\begin{figure}
  \begin{center}
    \includegraphics[width=0.9\textwidth]{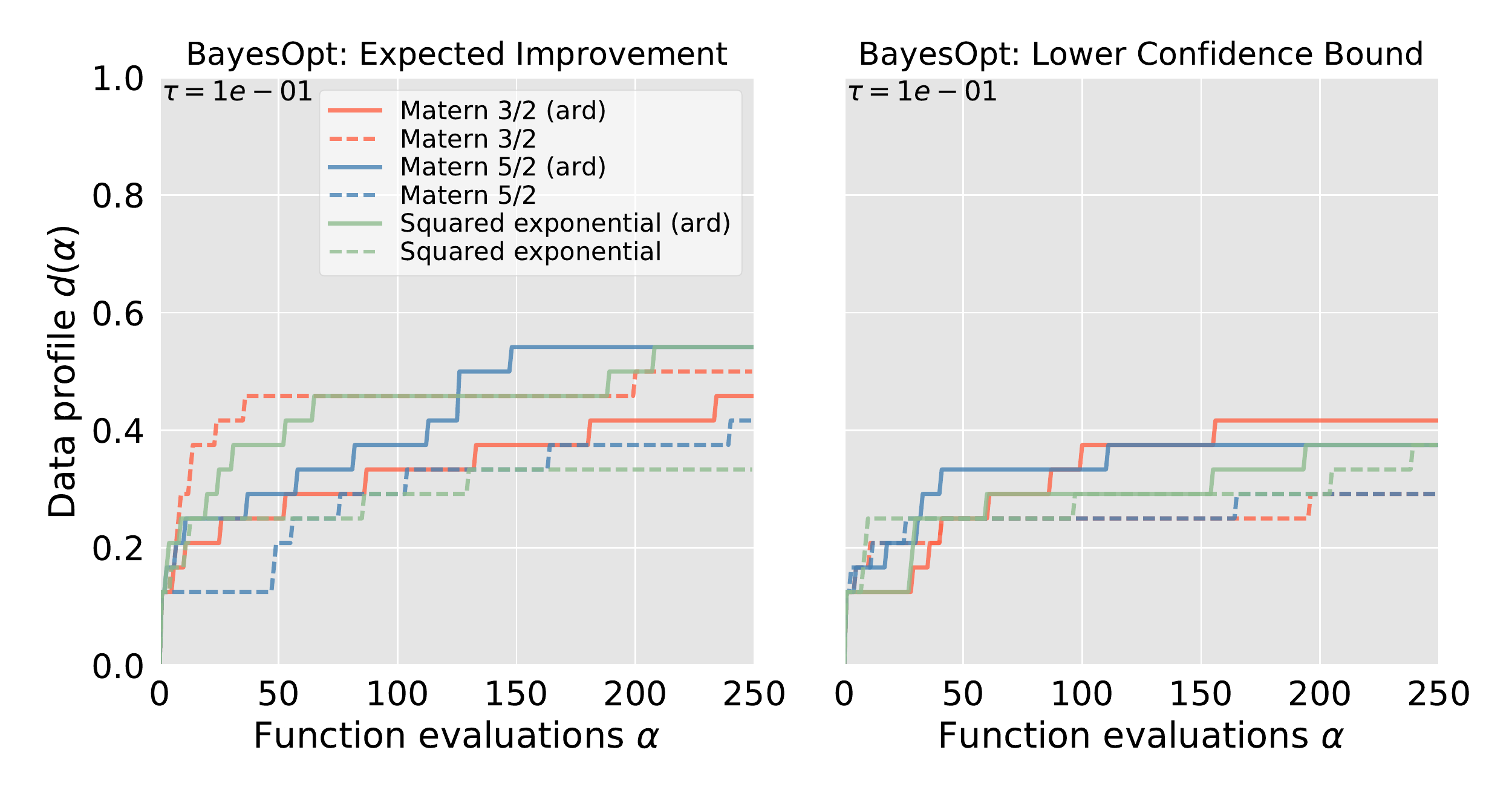}
    \includegraphics[width=0.9\textwidth]{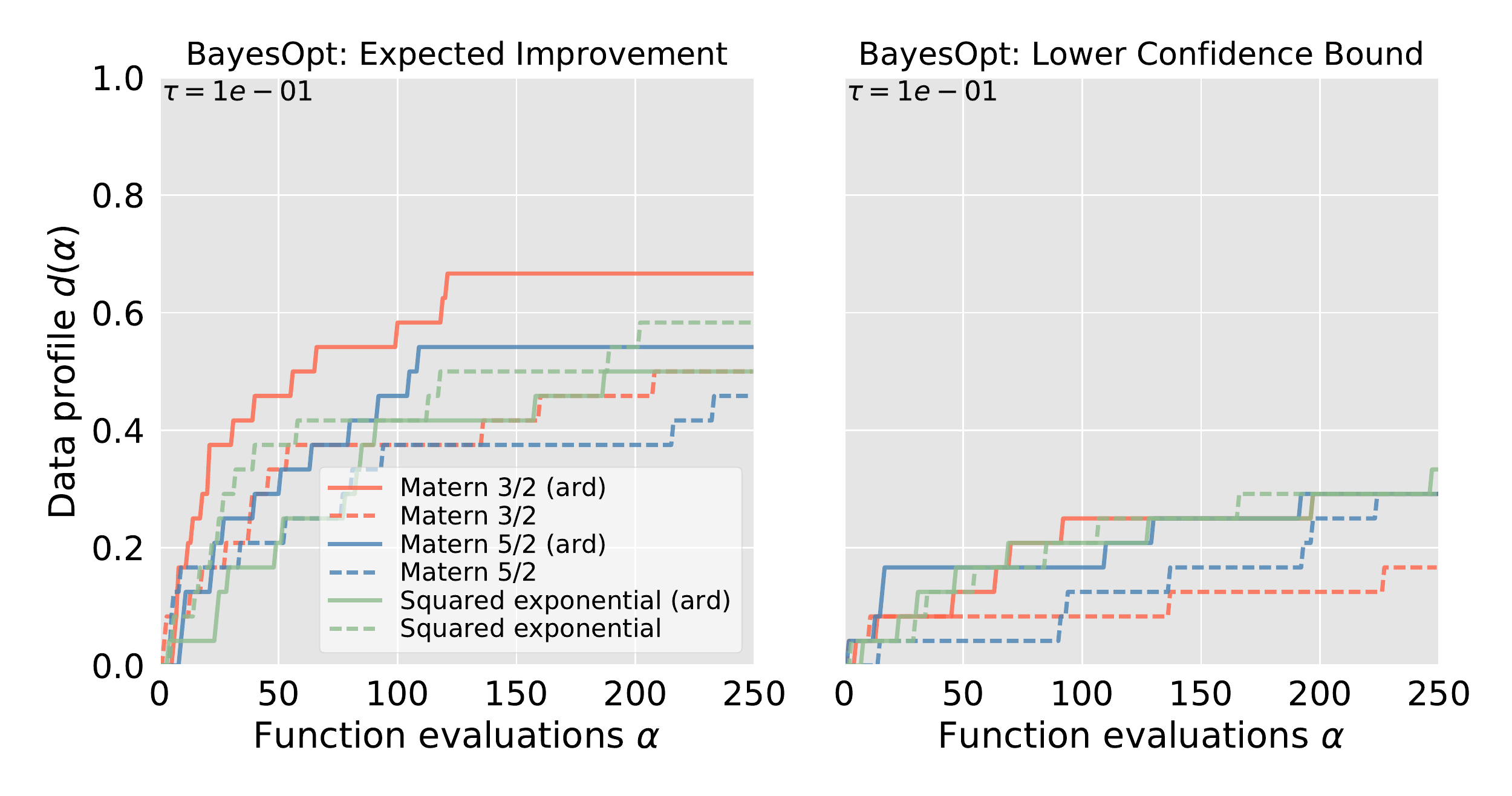}
    \includegraphics[width=0.9\textwidth]{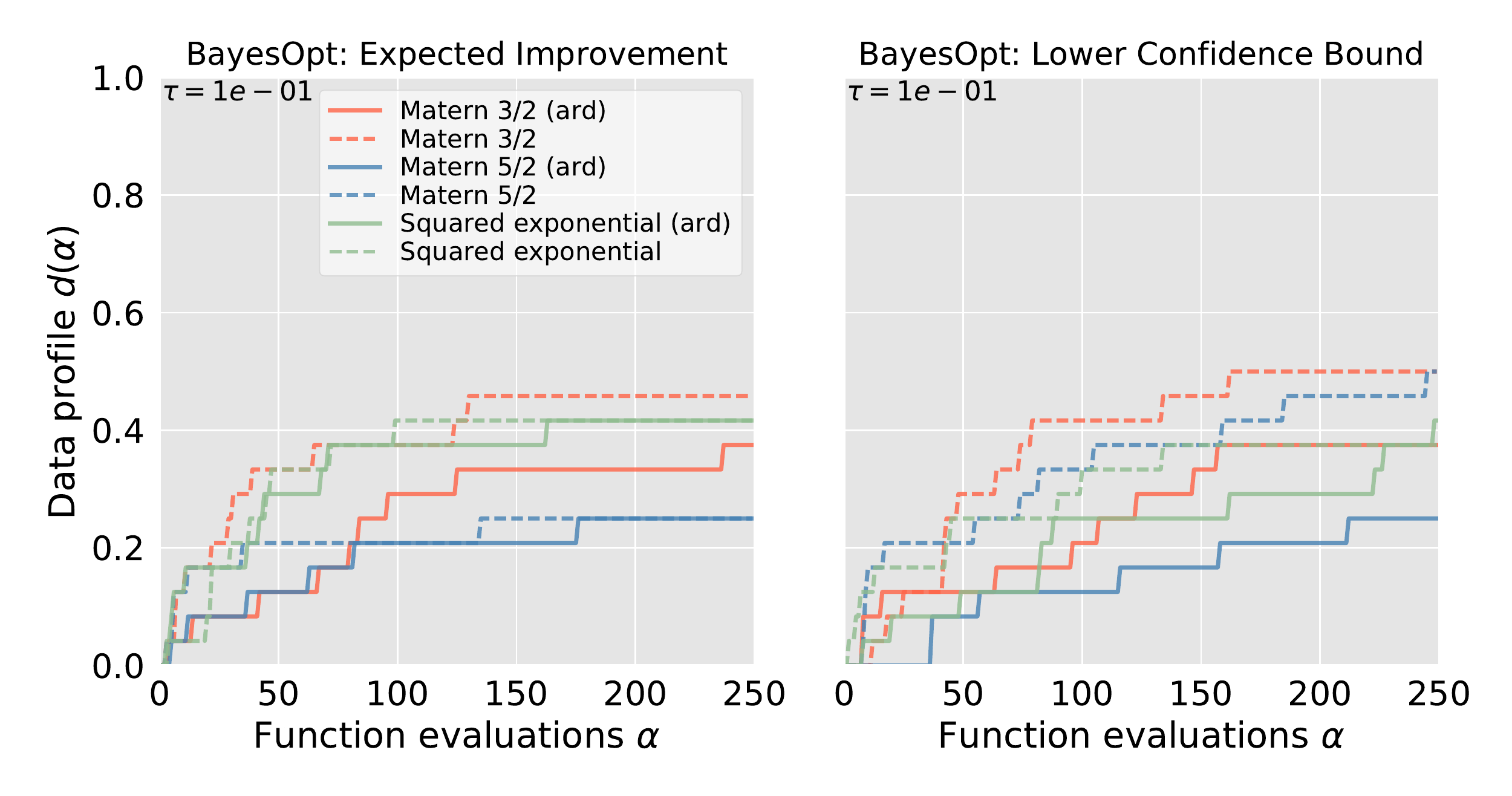}
    \caption{Separate data profiles for BayesOpt applied to
      three different parameter domains; \textit{small} (top row),
      \textit{medium} (middle row), and \textit{large} (bottom
      row).}
    \label{fig:prof_SML}
  \end{center}
\end{figure}
When we analyze the performance of BayesOpt in each of the three
parameter domains separately; \textit{small, medium, large}, see
Fig.~\ref{fig:prof_SML}, we note that the \textit{medium} and
\textit{large} domains stand out. In the former, the expected improvement acquisition
with the Matern 3/2 kernel and ARD algorithm achieves a 50\% performance
already after 50 iterations. This is the top performing BayesOpt
algorithm with the nuclear physics objective function. In the
\textit{small} and \textit{large} domains the Matern 3/2 kernel performs best
without ARD. Since BayesOpt will work best with a sensible prior, this
suggests rather short correlation lengths. In the \textit{large}
domain, the lower confidence-bound acquisition shows a tendency to perform better than
expected improvement. This is perhaps not too surprising since the parameter domain is
large enough to benefit from substantial exploration. Clearly, the ARD
kernels require much more data in larger spaces in order to prune
irrelevant features of the objective function.

\subsection{Comparison with the POUNDERs algorithm \label{sec:pounders}}

To facilitate a benchmark and further analysis of the strengths and
weaknesses of BayesOpt, we have also optimized the nuclear physics
objective function using the POUNDERs (Practical Optimization Using No
Derivatives for sums of Squares) optimization
algorithm~\cite{SWCHAP14}. This is a well-tested derivative-free
algorithm for optimizing non-linear least squares problems. Indeed, it
has been applied successfully in basic nuclear physics research since
almost a decade~\cite{Kortelainen2010,Ekstrom2013}. The key feature of
POUNDERs is that it assumes a least-squares structure of the objective
function, i.e. that it consists of a sum of squared residuals $R_i(x)$
written as
\begin{equation}
  f(\mathbf{x}) = \frac{1}{2}\sum_{i=1}^{p}R_{i}(\mathbf{x})^2.
  \label{eq:obj}
\end{equation}
Tailoring an optimization algorithm to exploit this mathematical
structure, i.e. that each term is a squared function of the parameters
$x$, is very fruitful. A quadratic model,
\begin{equation}
  m_k(\mathbf{x}_k+\mathbf{s}) = f(\mathbf{x}_k) + \mathbf{g}_k^T \mathbf{s} + \frac{1}{2}\mathbf{s}^T\mathbf{H}_k \mathbf{s}
\end{equation}
for an objective function $f$, at the current iterate $\mathbf{x}_k$,
with $\mathbf{g}_k = \nabla f(\mathbf{x}_k)$ and $\mathbf{H}_k =
\nabla^2 f(\mathbf{x}_k)$, is a very common choice. If the
corresponding derivatives are known the subproblem of minimizing $m_k$
can be solved quite efficiently. However, derivatives $\nabla f$ and
$\nabla^2 f$ are
considered unavailable for the present problems and only a set of
function values $f(\mathbf{y}^{(j)})$, and residual values
$R_{i}(\mathbf{y}^{(j)})$, for some set $\mathbf{Y} =
\left\{\mathbf{y}^{(1)},\mathbf{y}^{(2)},\ldots,\mathbf{y}^{(n)}\right
\}$ can be accessed. POUNDERs sets up a quadratic model for each
residual $i$ in Eq.~\ref{eq:obj}
\begin{equation}
  q^{(i)}_{k}(\mathbf{x}) = c^{(i)} +  (\mathbf{x} - \mathbf{x}^k)^T \mathbf{g}^{(i)} + \frac{1}{2}(\mathbf{x}-\mathbf{x}^k)^T\mathbf{H}^{(i)}(\mathbf{x}-\mathbf{x}^k)
\end{equation}
centered around the current iterate $\mathbf{x}_k \in \mathcal{X} \in
\mathbb{R}^{D}$. The model for each residual $i$ is defined by the
parameters $c^{(i)} \in \mathbb{R}$, $\mathbf{g}^{(i)} \in \mathbb{R}^{D}$, and
$\mathbf{H}^{(i)} \in \mathbb{R}^{D \times D}$. These
model parameters are determined by solving an interpolation problem in
$\mathbb{R}^D$, see Ref. ~\cite{SWCHAP14} and references therein. Once
the model parameters are obtained, they can be used to approximate the
derivatives of the objective function. In principle, the first- and
second-order derivatives of $f(\mathbf{x})$ with respect to $\mathbf{x}$ are given by
$\sum_i \nabla R_i(\mathbf{x}) R_i(\mathbf{x})$ and $\sum_i (\nabla R_i(\mathbf{x})\nabla R_i(\mathbf{x})
+ R_i(\mathbf{x})\nabla^2R_i(\mathbf{x}))$, respectively. Consequently, POUNDERs sets up
a master model $M_k$ for the objective function
\begin{equation}
  M_k(\mathbf{x}_k+\mathbf{s}) = f(\mathbf{x}_k) + \mathbf{s}^T \sum_{i=1}^{p} R_i(\mathbf{x}_k)\mathbf{g}^{(i)} + \frac{1}{2}\mathbf{s}^T \sum_{i=1}^{p} \left[ \mathbf{g}^{(i)}\left( \mathbf{g}^{(i)} \right)^T + R_i(\mathbf{x}_k) \mathbf{H}^{(i)}\right]\mathbf{s}.
\end{equation}
It should be noted that the master model itself does not interpolate the
objective over the interpolation set $\mathbf{Y}$.

POUNDERs is a trust-region method. The master model $M_k$ is believed
to approximate $f$ in a neighborhood $B_k$ of the current iterate
$\mathbf{x}_k$, where
\begin{equation}
  B_k = \left\{ \mathbf{x}\in\mathbb{R}^D: || \mathbf{x}-\mathbf{x}_k|| \leq \delta_k \right\}.
\end{equation}
The master model is therefore minimized over some $B_k$ with radius
$\delta_k >0$. The solution $\mathbf{s}_k$ to ${\rm min}\left\{
M_k(\mathbf{x}_k+\mathbf{s}): ||\mathbf{s}|| \leq \delta_k \right\}$
determines the next iterate $\mathbf{x}_{k+1}$ and a new trust region
radius $\delta_{k+1}$ is determined based on how good the model
prediction $M_k(\mathbf{x}_{k}+\mathbf{s}_k)$ was, see
Ref~\cite{SWCHAP14} for a full specification of the algorithm. We are
running POUNDERs in the default mode, meaning that we only have to
input $\mathbf{x}_{0} \in \mathbb{R}^{D}$, $M_0$, some initial
trust-region radius $0< \delta_k < 0.5$, and lower and upper bounds
$l_i, u_i$ of the domain $\mathcal{X}$. We also scale the problem such
that the bounds correspond to a unit hypercube $[0,1]^{D}$. This
scaling is not performed in BayesOpt.

\begin{figure}
  \begin{center}
    \includegraphics[width=0.9\textwidth]{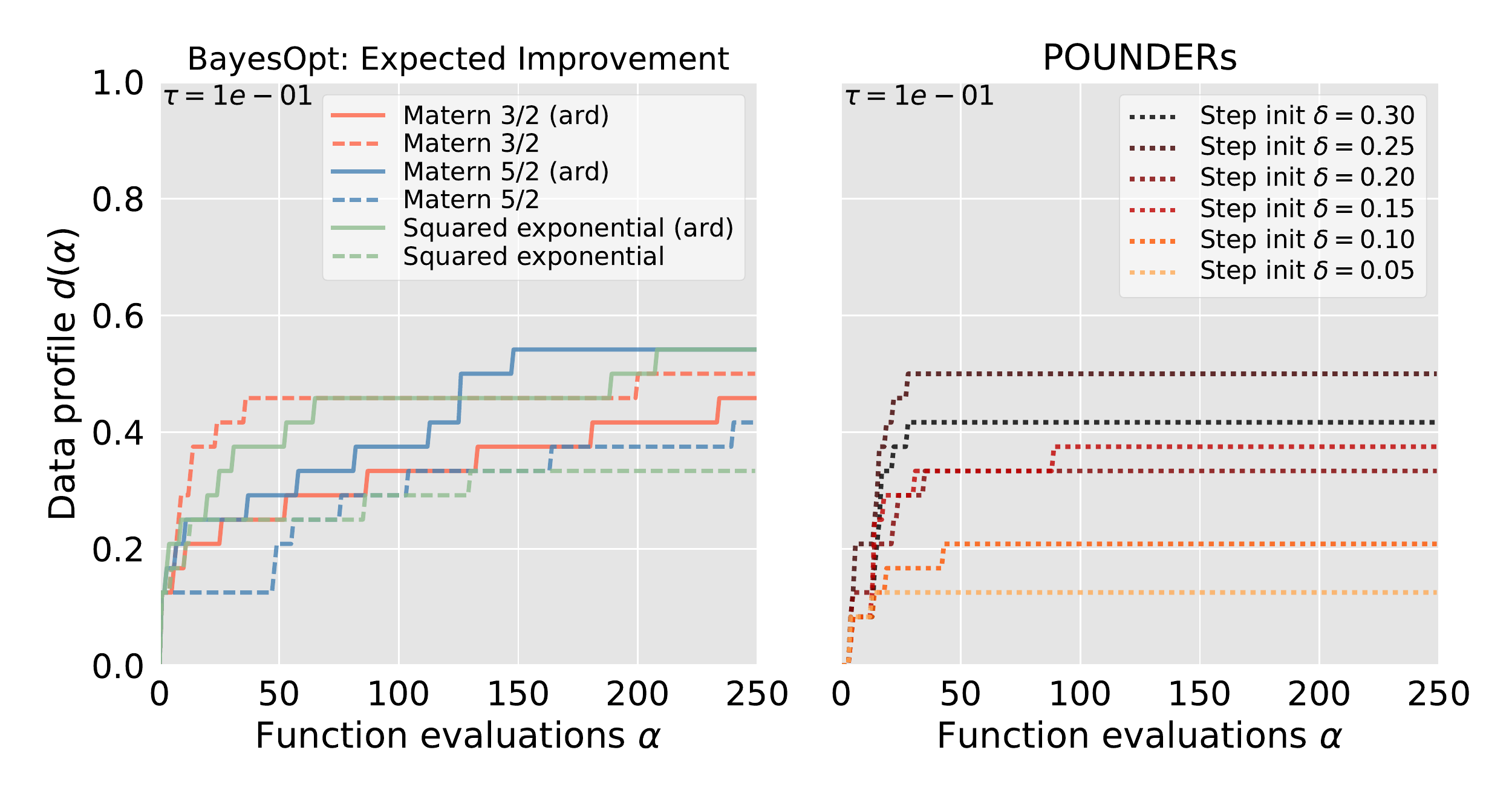}
    \includegraphics[width=0.9\textwidth]{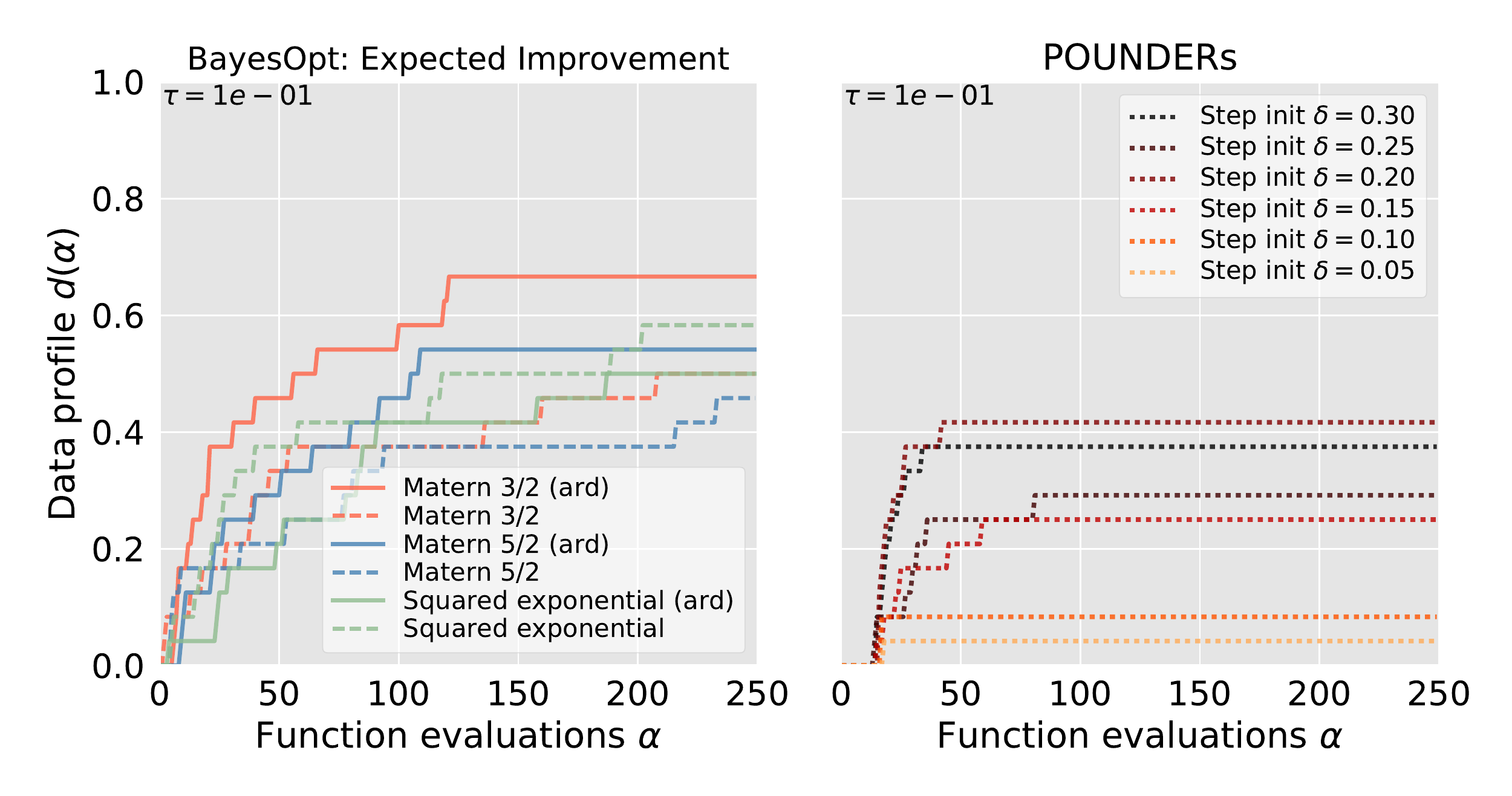}
    \includegraphics[width=0.9\textwidth]{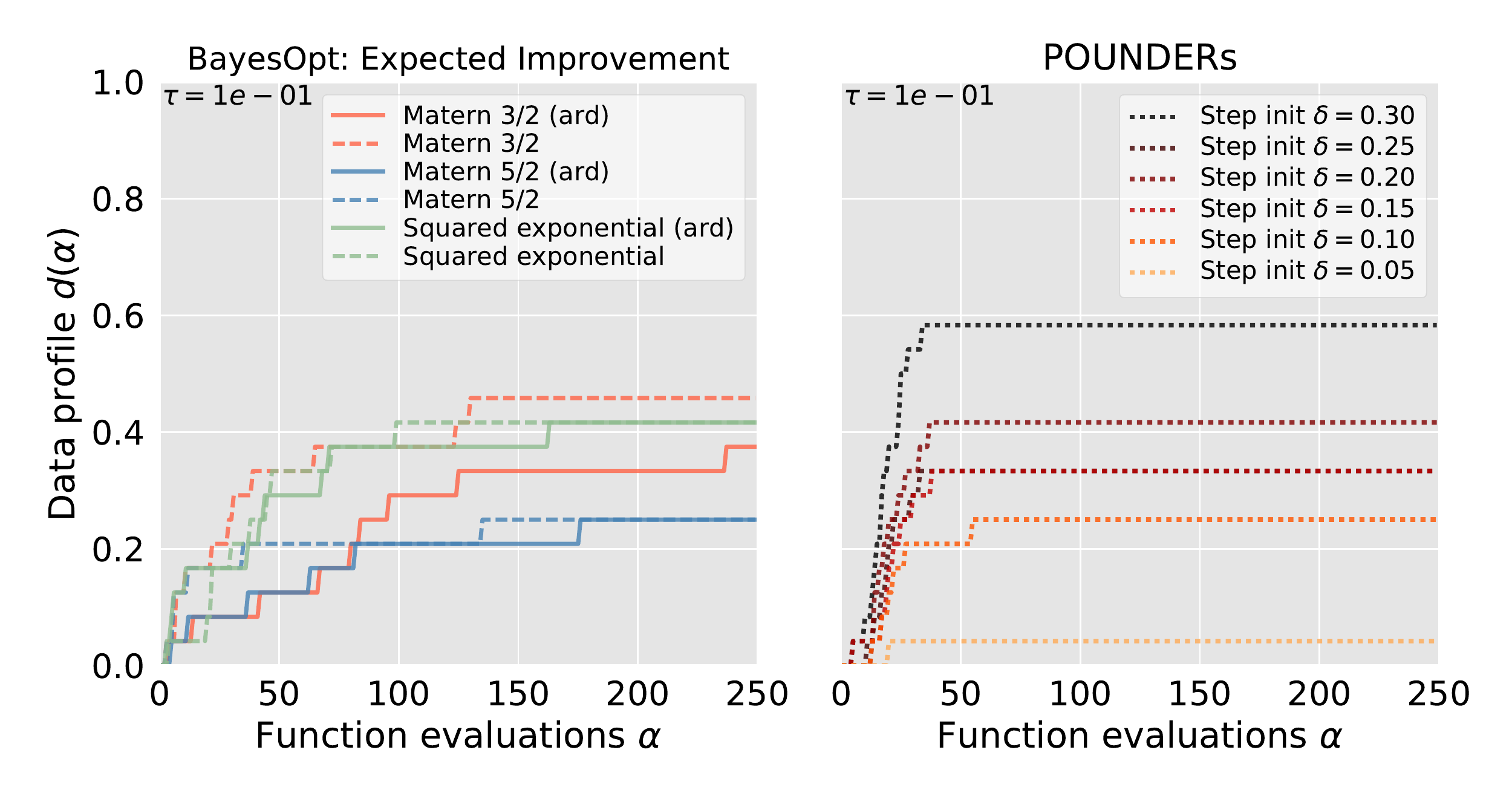}
    \caption{Separate data profiles for BayesOpt with the expected improvement
      acquisition function (left column) and POUNDERs (right column)
      applied to three different parameter domains; \textit{small}
      (top row), \textit{medium} (middle row), and \textit{large}
      (bottom row).}
    \label{fig:prof_bo_pounders_SML}
  \end{center}
\end{figure}

In Fig.~\ref{fig:prof_bo_pounders_SML} we present the data profiles
for POUNDERs applied to the physics objective function in the
\textit{small,medium}, and \textit{large} domains and compare with
BayesOpt. The results are only compared with the expected improvement acquisition
function as it was shown in Fig.~\ref{fig:prof_SML} to perform
significantly better than the lower confidence-bound acquisition function for this
optimization problem. We remind the reader that all algorithms are
initiated from an identical set of 24 starting points
$\{\mathbf{x}_1,\mathbf{x}_2,\ldots,\mathbf{x}_{24}\}$ for each
parameter domain.
As seen clearly in Fig.~\ref{fig:prof_bo_pounders_SML}, the choice of
initial trust-region radius $\delta_0$ determines the performance of
POUNDERs. Setting the initial radius too small
($\delta_0 \lesssim 0.15$) hampers the POUNDERs algorithm by trapping
it in a shallow local minimum. This is not an issue with BayesOpt
which continues to improve as more and more function values extends
the data vector $\mathcal{D}$.  As we have already noted several
times, the overall performance of BayesOpt depends crucially on the
prior. We also note that when POUNDERs performs well, it does so
within rather few function evaluations, but halts once it is trapped
in a local minimum. The advantages of BayesOpt are most prominent when
optimizing over the \textit{medium} domain. Regardless of kernel,
BayesOpt performs rather well even with few function evaluations. In
the \textit{large} domain, the good performance of POUNDERs clearly
indicates the usefulness of encoding prior knowledge about the
mathematical structure of the objective function. There is likely some
large scale structure in the objective function that the BayesOpt
kernel would benefit to learn about. Therefore, it would probably be
advantageous to amend the BayesOpt prior with a polynomial regression
term fitted to the first few evaluations of the objective function.

\begin{figure}
  \begin{center}
    \includegraphics[width=0.9\textwidth]{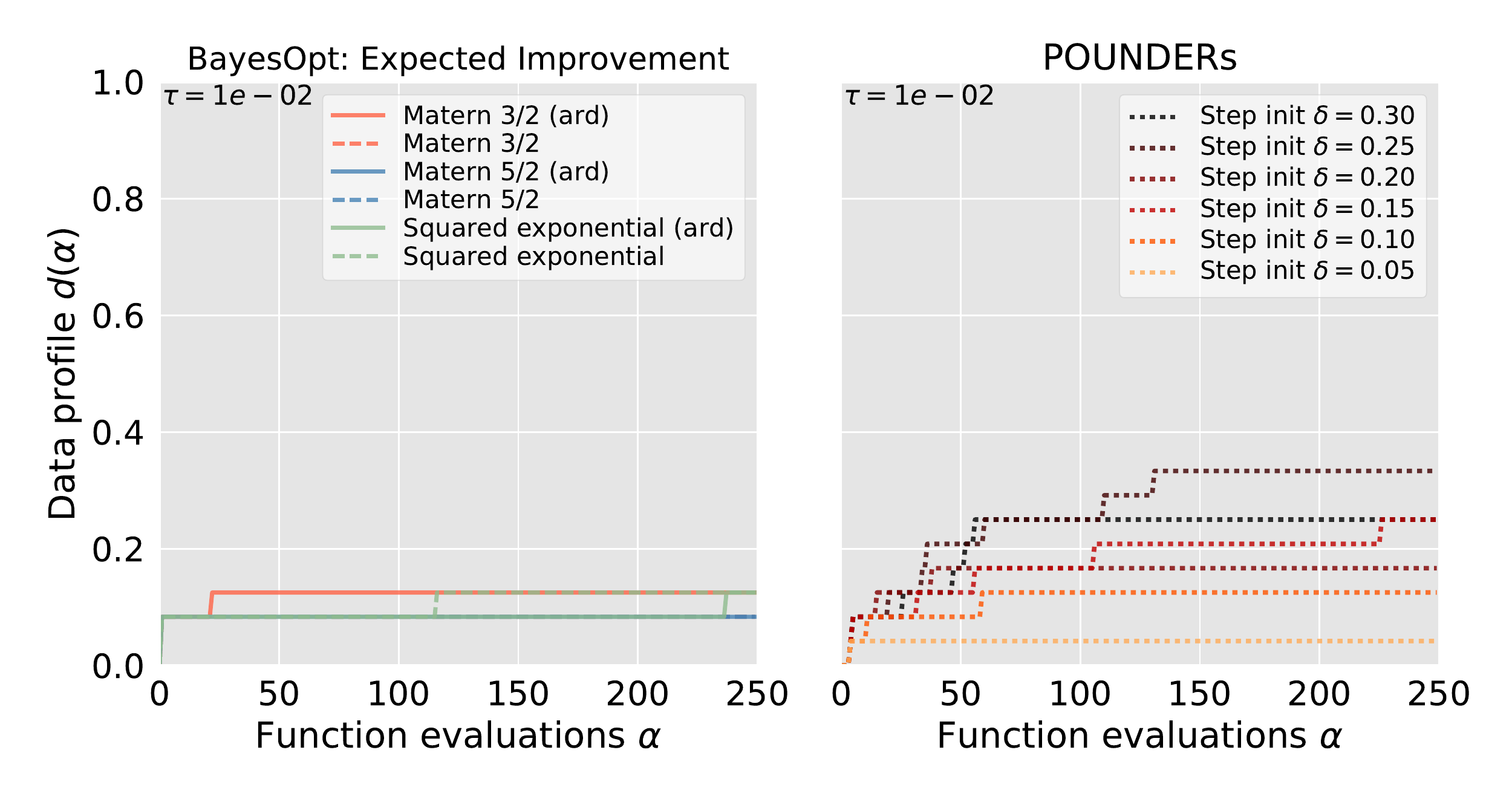}
    \caption{Data profiles comparing BayesOpt with the expected improvement
      acquisition function (left) and POUNDERs (right) for the nuclear
      physics objective function in the \textit{small} domain. Note
      that the convergence criterion is set to $\tau=0.01$. Only
      POUNDERs can achieve a reasonable level of performance.}
    \label{fig:prof_bo_pounders_S001}
  \end{center}
\end{figure}

BayesOpt is not intended for pinpointing the exact location of an
optimum. In the neighborhood of an optimum most objective functions
can be well approximated by a quadratic polynomial. For this reason,
POUNDERs will always outperform BayesOpt when decreasing $\tau$ in the
convergence criterion given in Eq.~\ref{eq:crit}. For $\tau=0.01$,
i.e. a convergence criterion corresponding to a 99\% reduction of the
objective function, BayesOpt will only reach a performance of
$d(\alpha<250)\approx 0.15$, whereas POUNDERs can approach $d(\alpha)
\approx 0.35$ for an optimal choice of the initial trust-region radius
$\delta_0$, see Fig.~\ref{fig:prof_bo_pounders_S001}.

\section{Summary and Outlook\label{sec:summary}}
Some of the most interesting optimization problems in nuclear
physics, as well as other fields of science, typically render
computationally expensive objective functions defined on
multi-dimensional parameter domains. Moreover, derivatives with
respect to those parameters are usually not accessible. 

In this paper we explore the potential benefits of BayesOpt
(Sec.~\ref{sec:bayesopt}) for efficiently exploring the parameter
space of a $\chi$EFT (\ref{app:nnlo}) in computationally
complex circumstances. A local minimum, with realistic physical
properties, in this parameter domain allows for numerical simulations
of atomic nuclei and therefore improves our understanding of the origin,
evolution, and structure of all matter in the universe. The underlying
optimization problem is therefore well known in the nuclear physics
research community and several classes of numerical optimization
algorithms have already been employed;
derivative-based~\cite{Carlsson:2016} as well as derivative-free
approaches~\cite{Ekstrom2013}.

BayesOpt presupposes a prior on the objective function, usually in the form of
a $\mathcal{GP}$. The original optimization challenge is transformed to a
design problem that boils down to choosing an appropriate acquisition
function to facilitate an exploration-exploitation balance. This
choice is encoded in a utility function that decides where to collect
training data for the $\mathcal{GP}$. Several choices of kernels and utility
functions exist. Our initial studies of BayesOpt applied to a set of
six test functions with parameter domains in $\mathbb{R}^{2,4,8}$
clearly demonstrate the importance of a sensible prior assumption of
the objective function, see Fig.~\ref{fig:prof_test_2d}. From this
analysis it is also clear that BayesOpt performs rather well in
low-dimensional ($\mathbb{R}^{2}$ and $\mathbb{R}^4$) parameter
domains. It turns out that the choice of acquisition function is even
more important than the choice of $\mathcal{GP}$-kernel, see
Fig.~\ref{fig:prof_Ackley}. This is something we see also when we
study the data profiles of BayesOpt applied to the nuclear
physics objective, see Fig.~\ref{fig:prof_SML}.

Our main findings and conclusions can be summarized as follows:

\begin{itemize}

\item
In general, BayesOpt will never find a narrow minimum nor be useful
for extracting the exact location of any optimum. For that to work, in
anything but a trivial case, it is necessary to have detailed
information about the objective function and successfully encode this
prior knowledge into the algorithm. This is a situation that we
typically do not have, since BayesOpt is applied to computationally
expensive objective functions. Instead, BayesOpt might find use as a
first stage in a hierarchical optimization protocol to identify an
interesting region of the parameter domain. It might also be
advantageous to design an acquisition function that is more
explorative during the first iterations, and then switch to an
acquisition function that exploits more than it explores.

\item
When we compare with the POUNDERs algorithm,
Sec.~\ref{sec:pounders}---a derivative-free optimization algorithm
that successfully incorporates the squared-sum structure of the
objective function---we find that BayesOpt in \textit{ab initio}
nuclear physics would probably benefit from a prior with a polynomial
regression term to efficiently capture the large scale structure of
the objective function.

\item
We find that the choice of acquisition function is more important than
the specific form of the $\mathcal{GP}$-kernel. For the present case,
the expected improvement acquisition function performed slightly
better than the lower confidence-bound in smaller parameter domains,
while more exploration as achieved with the lower confidence-bound
acquisition function, was shown to be beneficial in larger domains.

\item The $\mathcal{GP}$-kernel can be improved with ARD tuning of the
  hyperparameters. However, this feature is only useful if a minimum
  number of iterations can be afforded. In fact, the ARD kernels
  requires significantly more data in larger spaces in order to prune
  irrelevant features of the objective function.

\item Although we employ an objective function consisting of
  independent scattering cross sections, and all of them with similar
  and low computational cost, a multi-fidelity scenario is equally
  probable. Consider e.g.\ to calibrate an EFT model of the nuclear
  interaction using scattering cross sections \textit{and} data on
  bound states in multi-nucleon systems such as isotopes of oxygen and
  calcium. In such scenarios, where the computational cost of solving
  the Schr\"odinger equation for bound-states of a nucleus with $A$
  nucleons naively grow exponentially with $A$, it would be
  interesting to study the benefits of existing BayesOpt frameworks
  that can maximize information gathering across multiple functions
  with varying degrees of accuracy and computational cost. See
  e.g. Ref.~\cite{Yuxin2018} and references therein.

\end{itemize}

\section*{Acknowledgments}
The research leading to these result received financial support from
the BigData@Chalmers initiative. This research also received funding
from the European Research Council (ERC) under the European Union's 
Horizon 2020 research and innovation programme (grant agreement No
758027), the Swedish Research Council under Grants No. 2015-00225
and 2017-04234, and the Marie Sklodowska Curie Actions, Cofund, Project
INCA 600398. Computations were performed on resources provided by the
Swedish National Infrastructure for Computing (SNIC) at NSC, Link\"oping.

\appendix
\section{Test functions \label{app:testfunctions}}
In this appendix we provide the expressions for the six test functions
that we used for initial analysis of BayesOpt.
\begin{itemize}
\item Ackley: one hole on a semi-flat surface with shallow local minima.
  \begin{align}
    \begin{split}
      {}& f(\mathbf{x}) = -A \exp\left[ \frac{B}{d}\sum_{i=1}^{d} x_i^2\right] - \exp\left[ \frac{1}{d} \sum_{i=1}^{d} \cos(C x_i)\right] + A + e, \\
      {}& \text{where } A=20, B=\frac{1}{2}, \text{ and } C=2\pi.\\
      {}& \text{Domain: }x_{i}\in [-30,+30], \text{ for } i=1,\ldots,d.\\
      {}& \text{Global minimum: } \mathbf{x}_{\star} = (0,\ldots,0), \text{ and } f(\mathbf{x}_{\star}) = 0.
    \end{split}
  \end{align}
\item Deceptive: very challenging multivariate test function for which the
  total size of the region with local minima is $5^D-1$ times
  larger than the region with the global minimum. This function has
  $3^D-1$ local minima in $\mathbb{R}^D$.
  \begin{align}
    \begin{split}
      {}& f(\mathbf{x}) = -\left[ \frac{1}{d} \sum_{i=1}^{d} g_{i}(x_i) \right]^2, \\
      {}& \text{where }
      g_i(x_i) = 
      \begin{cases}
        -\frac{x_i}{\alpha_i} + \frac{4}{5} & \text{if }  0 \leq x_i < \frac{4}{5}  \alpha_i \\
        +5\frac{x_i}{\alpha_i} -4 & \text{if } \frac{4}{5}\alpha_i <  x_i \leq \alpha_i \\
        +5\frac{x_i-\alpha_i}{\alpha_i-1} +1 & \text{if } \alpha_i < x_i \leq \frac{1+4\alpha_i}{5} \\
        +\frac{x_i-1}{1-\alpha_i} + \frac{4}{5} & \text{if }\frac{1+4\alpha_i}{5} <  x_i \leq 1 \\
      \end{cases} \\
                  {}& \text{and } \alpha_i = \frac{i}{d+1}.\\
                  {}& \text{Domain: } x_{i}\in [0,1], \text{ for } i=1,\ldots,d.\\
                  {}& \text{Global minimum: } \mathbf{x}_{\star} = (\alpha_1,\ldots,\alpha_d), \text{ and } f(\mathbf{x}_{\star}) = -1.
    \end{split}
  \end{align}
\item Rastrigin: spherical function with cosine modulation to generate frequent local minima.
  \begin{align}
    \begin{split}
      {}& f(\mathbf{x}) = 10n + \sum_{i=1}^d \left[ x_i^2 - 10\cos(2\pi x_i)\right].\\
      {}& \text{Domain: }x_{i}\in [-5.12,5.12], \text{ for } i=1,\ldots,d.\\
      {}& \text{Global minimum: } \mathbf{x}_{\star} = (0,\ldots,0), \text{ and } f(\mathbf{x}_{\star}) = 0.
    \end{split}
  \end{align}
\item Rosenbrock: classic test function with minimum located in very shallow valley.
    \begin{align}
    \begin{split}
      {}& f(\mathbf{x}) = \sum_{i=1}^{d-1} \left[ 100(x_{i+1} - x_i^2)^2 + (1-x_i)^2\right].\\
      {}& \text{Domain: }x_{i}\in [-2.048,2.048], \text{ for } i=1,\ldots,d.\\
      {}& \text{Global minimum: } \mathbf{x}_{\star} = (1,\ldots,1), \text{ and } f(\mathbf{x}_{\star}) = 0.
    \end{split}
    \end{align}
  \item Schwefel: smooth surface with several local minima and a
    global minimum located far away in a corner, which in turn is far
    away from the second best local minimum.
    \begin{align}
    \begin{split}
      {}& f(\mathbf{x}) = 418.9829d - \sum_{i=1}^{d} \left[ x_i -\sin\left( \sqrt{|x_i|}\right) \right].\\
      {}& \text{Domain: }x_{i}\in [-500,500], \text{ for } i=1,\ldots,d.\\
      {}& \text{Global minimum: } \mathbf{x}_{\star} = (420.9687,\ldots,420.9687), \text{ and } f(\mathbf{x}_{\star}) \approx 0.
    \end{split}
  \end{align}
  \item Sphere: in many ways the simplest possible test function. It is convex and unimodal.
    \begin{align}
      \begin{split}
        {}& f(\mathbf{x}) = \sum_{i=1}^{d} x_i^2. \\
        {}& \text{Domain: }x_{i}\in [-5.12,5.12], \text{ for } i=1,\ldots,d.\\
        {}& \text{Global minimum: } \mathbf{x}_{\star} = (0,\ldots,0), \text{ and } f(\mathbf{x}_{\star}) = 0.
      \end{split}
    \end{align}
\end{itemize}

\section{Physics model: $\chi$EFT for neutron-proton scattering at NNLO \label{app:nnlo}}
In the interest of keeping this paper somewhat self-contained, we
briefly review the formalism that underpins the physics model we are
optimizing. To that end, we include a selected set of expressions that
are representative for the full theoretical framework for computing
neutron-proton ($np$) scattering cross section starting from a
potential description in $\chi$EFT. Exhaustive reviews of EFT are given
in~\cite{Bedaque:2002gm, EHM09, 0000000102130473}. The model
evaluation $\mathcal{O}^{\rm model}(\mathbf{x})$ of an $np$ scattering
cross section to be confronted with an experimentally determined value
$\mathcal{O}^{\rm experiment}$ at a given on-shell momentum $k$
proceeds in three steps:
\begin{enumerate}
\item compute the momentum-space proton-neutron interaction potential.
\item use the potential to compute the quantum-mechanical scattering
  amplitude by solving the non-relativistic Lippmann-Schwinger
  equation.
\item use the amplitude to compute a model value for the scattering
  cross section by evaluating the spin-scattering matrix.
\end{enumerate}

\subsection{The $np$ interaction potential at NNLO in $\chi$EFT}
\begin{figure}
  \begin{center}
    \includegraphics[width=1.0\linewidth]{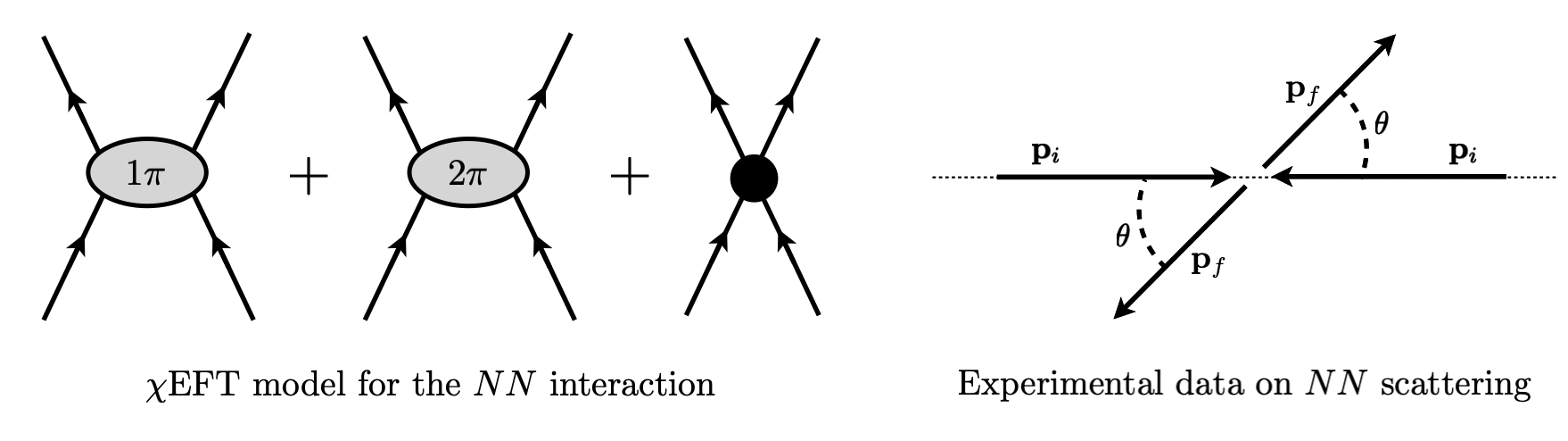}
    \caption{$NN$ scattering (right), and associated experimental
      data, can be used to determine the values of the parameters
      (LECs) of the $\chi$EFT NNLO model (left).}
    \label{fig:model}
  \end{center}
\end{figure}
The interaction potential in $\chi$EFT consists of non-polynomial
terms that describe the long-range part corresponding to pion
exchanges, and contact-type
interactions given by polynomial expressions corresponding to a short-range part:
\begin{equation}
  V(\mathbf{p_f},\mathbf{p_i}) =V_{\rm long-range}(\mathbf{p_f},\mathbf{p_i}) + V_{\rm contact}(\mathbf{p_f},\mathbf{p_i}),
\end{equation}
where $\mathbf{p_f}$ and $\mathbf{p_i}$ denote the final and initial
nucleon momenta in the center-of-mass system (CMS), and the special
case $|\mathbf{p_f}\,| = |\mathbf{p_i}| = k$ corresponds to an on-shell momentum.
The potential can be written as a systematic expansion with
high-order terms being less important than low-order ones.
In this work we employ a potential including terms up to NNLO in
$\chi$EFT. This means that there are terms also at leading-order (LO)
and next-to-leading order (NLO). In general, at higher orders there
are more pion exchanges and higher-order polynomial terms in momenta
that flow through the contact diagrams. At LO, the one-pion exchange
potential is given by
\begin{equation}
  V_{1\pi}^{(\rm LO)}(\mathbf{p_f},\mathbf{p_i}) = \frac{g_A^2}{4f_\pi^2}
\: 
\boldtau_1 \cdot \boldtau_2 
\:
\frac{
\boldsigma_1 \cdot \mathbf{q} \,\, \boldsigma_2 \cdot \mathbf{q}}
{q^2 + m_\pi^2} 
\,,
\end{equation}
where $\mathbf{q} \equiv \mathbf{p_f} - \mathbf{p_i}$ is the momentum transfer,
$\boldsigma_{1,2}$ and $\boldtau_{1,2}$ are the spin and isospin
operators of nucleon 1 and 2, $g_A$, $f_\pi$, and $m_\pi$ denote the
axial-vector coupling constant, the pion decay constant, and the pion
mass, respectively. We use $f_\pi=92.4$ MeV and $g_A=1.29$ throughout
this work. Higher order corrections to one-pion exchange renormalize
the coupling constants $g_A$ and $f_{\pi}$, and the LO long-ranged
part is considered parameter-free in this work. Up to NNLO, leading
and sub-leading two-pion exchange enters and the long-ranged part of
the interaction is given by
\begin{align}
  \label{eq:nnlo2pi}
  \begin{split}
  {}& V_{2\pi}^{({\rm NLO})}(\mathbf{p_f},\mathbf{p_i}) + V_{2\pi}^{({\rm NNLO})}(\mathbf{p_f},\mathbf{p_i}) = \\
  {}& L(q;\tilde{\Lambda})\left[ (\boldtau_1 \cdot \boldtau_2 )\left\{ 4m_{\pi}^2(1+4g_A^2-5g_A^4) +  
    q^2(1+10g_A^2 -23g_A^4) - \frac{48g_A^4m_{\pi}^4}{w^2} \right\}\right. - \\
    {}& \left. 18g_A^4((\boldsigma_1 \cdot \mathbf{q} \,\, \boldsigma_2 \cdot \mathbf{q}) -q^2 (\boldsigma_1 \cdot \boldsigma_2)) \right] + 
  A(q;\tilde{\Lambda})\left[(2m_{\pi}^2 + q^2)(2m_{\pi}^2(c_3-2c_1) + c_3q^2) \right. - \\
    {}& \frac{1}{6}c_4w^2 ((\boldtau_1\cdot\boldtau_2)\,(\boldsigma_1 \cdot \mathbf{q} \,\, \boldsigma_2 \cdot \mathbf{q}) -
    \left. q^2 (\boldtau_1\cdot\boldtau_2)(\boldsigma_1 \cdot \boldsigma_2))\right]
  \end{split}
\end{align}
where $w^2 = (4m_{\pi}^2 + q^2)$ and the loop functions are here written as
\begin{align}
  \begin{split}
    {}& L(q;\tilde{\Lambda}) = \frac{w}{768 \pi^2 f_{\pi}^2 q} {\rm ln} \frac{\tilde{\Lambda}^2(2m_{\pi}^2 + q^2) - 2m_{\pi}^2q^2 + \tilde{\Lambda}\sqrt{\tilde{\Lambda}^2-4m_{\pi}^2}qw}{2m_{\pi}^2(\tilde{\Lambda}^2 + q^2)}\\
    {}& A(q;\tilde{\Lambda}) = \frac{3g_A^2}{32 \pi f_{\pi}^4q}{\rm arctan}\frac{q(\tilde{\Lambda} - 2m_{\pi})}{q^2+2\tilde{\Lambda}m_{\pi}}
  \end{split}
\end{align}
and the so-called spectral function cutoff is set to
$\tilde{\Lambda}=700$ MeV. The long-ranged part contains three of 12
unknown LECs that we seek to constrain using BayesOpt, denoted $c_1$,
$c_3$, and $c_4$ in Eq.~\eqref{eq:nnlo2pi}. The remaining nine
LECs control the short-ranged part of the NNLO potential, which can be
written as a linear combination of terms polynomial in the initial and
final momenta
\begin{align}
  \begin{split}
    V_{\rm contact}(\mathbf{p_f},\mathbf{p_i}) = {}& C_S + C_T(\boldsigma_1 \cdot \boldsigma_2) + C_1q^2 + C_2k^2 + (C_3q^2 + C_4k^2)(\boldsigma_1 \cdot \boldsigma_2) + \\
    {}& C_5(-i\mathbf{S}\cdot(\mathbf{q} \times \mathbf{k})) + C_6(\boldsigma_1 \cdot \mathbf{q})(\boldsigma_2 \cdot \mathbf{q}) + C_7(\boldsigma_1 \cdot \mathbf{k})(\boldsigma_2 \cdot \mathbf{k})
  \end{split}
\end{align}

\subsection{Proton-neutron scattering observables}
Proton-neutron elastic scattering observables are calculated from the
spin-scattering matrix $M$~\cite{bystricky1978,lafrance1980}. This is
a $4\times4$ matrix in spin-space that operates on the initial state
to give the scattered part of the final state. $M$ is related to the
conventional scattering matrix $S$ by $M = \frac{2\pi}{ik} (S-1)$,
where $k$ is the relative momentum between the nucleons. The
$S$-matrix for the scattering channel with angular momentum $J$ can be
parameterized by the Stapp phase shifts
$S=e^{2i\delta_J}$~\cite{stapp1957}. The Stapp phase shifts are
calculated from the potential $V(\mathbf{p_f},\mathbf{p_i})$ by
solving the Lippmann-Schwinger equation. This equation describes
quantum-mechanical scattering, and is an integral equation of the
Fredholm type that can be solved as a matrix equation. In our
application, and for each value of the on-shell momentum $k$, this
amounts to inverting a 200-by-200 matrix followed by a matrix-vector
multiplication. The matrix inversion prevents linearizing this
particular EFT model in its parameters. Although, the matrix inverse
is not particularly time-consuming in the present case, it should be
pointed out that the complexity of the corresponding
quantum-mechanical equations for describing scattering states, or
bound states, of more than two nucleons typically scale exponentially
with the particle number.

\section*{References}
\bibliography{bonp}
\bibliographystyle{unsrt}
\end{document}